\documentclass[12pt]{article}
\usepackage{epsfig}
\usepackage{amsmath}
\usepackage{hhline}
\usepackage{amssymb}
\usepackage{times}
\usepackage{cite}
\newlength{\dinwidth}
\newlength{\dinmargin}
\begin{document}  
%%%%%%%%%%%%%%%% Pre-defined commands, you can use for the most obvious notations
\newcommand{\pom}{{I\!\!P}}
\newcommand{\reg}{{I\!\!R}}
\newcommand{\slowpi}{\pi_{\mathit{slow}}}
\newcommand{\fiidiii}{F_2^{D(3)}}
\newcommand{\fiidiiiarg}{\fiidiii\,(\beta,\,Q^2,\,x)}
\newcommand{\n}{1.19\pm 0.06 (stat.) \pm0.07 (syst.)}
\newcommand{\nz}{1.30\pm 0.08 (stat.)^{+0.08}_{-0.14} (syst.)}
\newcommand{\fiidiiiful}{F_2^{D(4)}\,(\beta,\,Q^2,\,x,\,t)}
\newcommand{\fiipom}{\tilde F_2^D}
\newcommand{\ALPHA}{1.10\pm0.03 (stat.) \pm0.04 (syst.)}
\newcommand{\ALPHAZ}{1.15\pm0.04 (stat.)^{+0.04}_{-0.07} (syst.)}
\newcommand{\fiipomarg}{\fiipom\,(\beta,\,Q^2)}
\newcommand{\pomflux}{f_{\pom / p}}
\newcommand{\nxpom}{1.19\pm 0.06 (stat.) \pm0.07 (syst.)}
\newcommand {\gapprox}
   {\raisebox{-0.7ex}{$\stackrel {\textstyle>}{\sim}$}}
\newcommand {\lapprox}
   {\raisebox{-0.7ex}{$\stackrel {\textstyle<}{\sim}$}}
\def\gsim{\,\lower.25ex\hbox{$\scriptstyle\sim$}\kern-1.30ex%
\raise 0.55ex\hbox{$\scriptstyle >$}\,}
\def\lsim{\,\lower.25ex\hbox{$\scriptstyle\sim$}\kern-1.30ex%
\raise 0.55ex\hbox{$\scriptstyle <$}\,}
\newcommand{\pomfluxarg}{f_{\pom / p}\,(x_\pom)}
\newcommand{\dsf}{\mbox{$F_2^{D(3)}$}}
\newcommand{\dsfva}{\mbox{$F_2^{D(3)}(\beta,Q^2,x_{I\!\!P})$}}
\newcommand{\dsfvb}{\mbox{$F_2^{D(3)}(\beta,Q^2,x)$}}
\newcommand{\dsfpom}{$F_2^{I\!\!P}$}
\newcommand{\gap}{\stackrel{>}{\sim}}
\newcommand{\lap}{\stackrel{<}{\sim}}
\newcommand{\fem}{$F_2^{em}$}
\newcommand{\tsnmp}{$\tilde{\sigma}_{NC}(e^{\mp})$}
\newcommand{\tsnm}{$\tilde{\sigma}_{NC}(e^-)$}
\newcommand{\tsnp}{$\tilde{\sigma}_{NC}(e^+)$}
\newcommand{\st}{$\star$}
\newcommand{\sst}{$\star \star$}
\newcommand{\ssst}{$\star \star \star$}
\newcommand{\sssst}{$\star \star \star \star$}
\newcommand{\tw}{\theta_W}
\newcommand{\sw}{\sin{\theta_W}}
\newcommand{\cw}{\cos{\theta_W}}
\newcommand{\sww}{\sin^2{\theta_W}}
\newcommand{\cww}{\cos^2{\theta_W}}
\newcommand{\trm}{m_{\perp}}
\newcommand{\trp}{p_{\perp}}
\newcommand{\trmm}{m_{\perp}^2}
\newcommand{\trpp}{p_{\perp}^2}
\newcommand{\alp}{\alpha_s}

\newcommand{\alps}{\alpha_s}
\newcommand{\sqrts}{$\sqrt{s}$}
\newcommand{\LO}{$O(\alpha_s^0)$}
\newcommand{\Oa}{$O(\alpha_s)$}
\newcommand{\Oaa}{$O(\alpha_s^2)$}
\newcommand{\PT}{p_{\perp}}
\newcommand{\JPSI}{J/\psi}
\newcommand{\sh}{\hat{s}}
\newcommand{\uh}{\hat{u}}
\newcommand{\MP}{m_{J/\psi}}
\newcommand{\PO}{I\!\!P}
\newcommand{\xbj}{x}
\newcommand{\xpom}{x_{\PO}}
\newcommand{\ttbs}{\char'134}
\newcommand{\xpomlo}{3\times10^{-4}}  
\newcommand{\xpomup}{0.05}  
\newcommand{\dgr}{^\circ}
\newcommand{\pbarnt}{\,\mbox{{\rm pb$^{-1}$}}}
\newcommand{\gev}{\,\mbox{GeV}}
\newcommand{\WBoson}{\mbox{$W$}}
\newcommand{\fbarn}{\,\mbox{{\rm fb}}}
\newcommand{\fbarnt}{\,\mbox{{\rm fb$^{-1}$}}}
\newcommand{\dsdx}[1]{$d\sigma\!/\!d #1\,$}
\newcommand{\eV}{\mbox{e\hspace{-0.08em}V}}
%
% Some useful tex commands
%
\newcommand{\qsq}{\ensuremath{Q^2} }
\newcommand{\gevsq}{\ensuremath{\mathrm{GeV}^2} }
\newcommand{\et}{\ensuremath{E_t^*} }
\newcommand{\rap}{\ensuremath{\eta^*} }
\newcommand{\gp}{\ensuremath{\gamma^*}p }
\newcommand{\dsiget}{\ensuremath{{\rm d}\sigma_{ep}/{\rm d}E_t^*} }
\newcommand{\dsigrap}{\ensuremath{{\rm d}\sigma_{ep}/{\rm d}\eta^*} }

%%% Dstar stuff
\newcommand{\dstar}{\ensuremath{D^*}}
\newcommand{\dstarp}{\ensuremath{D^{*+}}}
\newcommand{\dstarm}{\ensuremath{D^{*-}}}
\newcommand{\dstarpm}{\ensuremath{D^{*\pm}}}
\newcommand{\zDs}{\ensuremath{z(\dstar )}}
\newcommand{\Wgp}{\ensuremath{W_{\gamma p}}}
\newcommand{\ptds}{\ensuremath{p_t(\dstar )}}
\newcommand{\etads}{\ensuremath{\eta(\dstar )}}
\newcommand{\ptj}{\ensuremath{p_t(\mbox{jet})}}
\newcommand{\ptjn}[1]{\ensuremath{p_t(\mbox{jet$_{#1}$})}}
\newcommand{\etaj}{\ensuremath{\eta(\mbox{jet})}}
\newcommand{\detadsj}{\ensuremath{\eta(\dstar )\, \mbox{-}\, \etaj}}

% Journal macro
\def\Journal#1#2#3#4{{#1} {\bf #2} (#3) #4}
\def\NCA{\em Nuovo Cimento}
\def\NIM{\em Nucl. Instrum. Methods}
\def\NIMA{{\em Nucl. Instrum. Methods} {\bf A}}
\def\NPB{{\em Nucl. Phys.}   {\bf B}}
\def\PLB{{\em Phys. Lett.}   {\bf B}}
\def\PRL{\em Phys. Rev. Lett.}
\def\PRD{{\em Phys. Rev.}    {\bf D}}
\def\ZPC{{\em Z. Phys.}      {\bf C}}
\def\EJC{{\em Eur. Phys. J.} {\bf C}}
\def\CPC{\em Comp. Phys. Commun.}

%%%%%%%%%%%%% my new commands %%%%%%%%%%%%%%%%
\newcommand{\beq}{\begin{equation}} 
\newcommand{\eeq}{\end{equation}} 
\newcommand{\ba}{\begin{eqnarray}} 
\newcommand{\ea}{\end{eqnarray}}

\newcommand{\Nobs}{N_{\rm{obs}}}
\newcommand{\Mall}{M_{\rm{all}}}
\newcommand{\pmin}{p_{\rm{\min}}}
\newcommand{\SPT} {\sum P_T}
\newcommand{\CosTh} {\cos \theta^*_{\rm{lead}}}
\newcommand{\Xlead} {X_{\rm{lead}}}

\newcommand{\ee}{\mbox{$e$-$e$}}
\newcommand{\emu}{\mbox{$e$-$\mu$}}
\newcommand{\ej}{\mbox{$e$-$j$}}
\newcommand{\enp}{\mbox{$e$-$\nu$}}
\newcommand{\epho}{\mbox{$\gamma$-$e$}}
\newcommand{\mumu}{\mbox{$\mu$-$\mu$}}
\newcommand{\mumumu}{\mbox{$\mu$-$\mu$-$\mu$}}
\newcommand{\muj}{\mbox{$\mu$-$j$}}
\newcommand{\mujj}{\mbox{$\mu$-$j$-$j$}}
\newcommand{\munp}{\mbox{$\mu$-$\nu$}}
\newcommand{\mupho}{\mbox{$\gamma$-$\mu$}}
\newcommand{\jj}{\mbox{$j$-$j$}}
\newcommand{\jnp}{\mbox{$\nu$-$j$}}
\newcommand{\jpho}{\mbox{$\gamma$-$j$}}
\newcommand{\nppho}{\mbox{$\gamma$-$\nu$}}
\newcommand{\phopho}{\mbox{$\gamma$-$\gamma$}}
\newcommand{\phophopho}{\mbox{$\gamma$-$\gamma$-$\gamma$}}
\newcommand{\eee}{\mbox{$e$-$e$-$e$}}
\newcommand{\eej}{\mbox{$e$-$e$-$j$}}
\newcommand{\eenp}{\mbox{$e$-$e$-$\nu$}}
\newcommand{\eepho}{\mbox{$\gamma$-$e$-$e$}}
\newcommand{\eemu}{\mbox{$e$-$e$-$\mu$}}
\newcommand{\emumu}{\mbox{$e$-$\mu$-$\mu$}}
\newcommand{\emunu}{\mbox{$e$-$\mu$-$\nu$}}
\newcommand{\jmumu}{\mbox{$\mu$-$\mu$-$j$}}
\newcommand{\mumunp}{\mbox{$\mu$-$\mu$-$\nu$}}
\newcommand{\mumuj}{\mbox{$\mu$-$\mu$-$j$}}
\newcommand{\mumupho}{\mbox{$\gamma$-$\mu$-$\mu$}}
\newcommand{\emuj}{\mbox{$e$-$\mu$-$j$}}
\newcommand{\emupho}{\mbox{$\gamma$-$e$-$\mu$}}
\newcommand{\emunp}{\mbox{$e$-$\mu$-$\nu$}}
\newcommand{\enppho}{\mbox{$\gamma$-$e$-$\nu$}}
\newcommand{\ejj}{\mbox{$e$-$j$-$j$}}
\newcommand{\ejnp}{\mbox{$e$-$\nu$-$j$}}
\newcommand{\ejpho}{\mbox{$\gamma$-$e$-$j$}}
\newcommand{\ephopho}{\mbox{$\gamma$-$\gamma$-$e$}}
\newcommand{\jphopho}{\mbox{$\gamma$-$\gamma$-$j$}}
\newcommand{\muphopho}{\mbox{$\gamma$-$\gamma$-$\mu$}}
\newcommand{\mujnp}{\mbox{$\mu$-$\nu$-$j$}}
\newcommand{\munppho}{\mbox{$\gamma$-$\mu$-$\nu$}}
\newcommand{\mujpho}{\mbox{$\gamma$-$\mu$-$j$}}
\newcommand{\jjj}{\mbox{$j$-$j$-$j$}}
\newcommand{\jjnp}{\mbox{$\nu$-$j$-$j$}}
\newcommand{\jjpho}{\mbox{$\gamma$-$j$-$j$}}
\newcommand{\jnppho}{\mbox{$\gamma$-$\nu$-$j$}}
\newcommand{\jjnppho}{\mbox{$\gamma$-$\nu$-$j$-$j$}}
\newcommand{\ejjj}{\mbox{$e$-$j$-$j$-$j$}}
\newcommand{\jjjj}{\mbox{$j$-$j$-$j$-$j$}}
\newcommand{\jjjnp}{\mbox{$\nu$-$j$-$j$-$j$}}
\newcommand{\jjjpho}{\mbox{$\gamma$-$j$-$j$-$j$}}
\newcommand{\ejjjj}{\mbox{$e$-$j$-$j$-$j$-$j$}}
\newcommand{\jjjjj}{\mbox{$j$-$j$-$j$-$j$-$j$}}
\newcommand{\jjjjnp}{\mbox{$\nu$-$j$-$j$-$j$-$j$}}
\newcommand{\ejjnp}{\mbox{$e$-$\nu$-$j$-$j$}}
\newcommand{\ejjpho}{\mbox{$\gamma$-$e$-$j$-$j$}}
\newcommand{\eejnp}{\mbox{$e$-$e$-$\nu$-$j$}}
\newcommand{\emujnp}{\mbox{$e$-$\mu$-$\nu$-$j$}}

\begin{titlepage}

\noindent
\begin{flushleft}
{\tt DESY 08-173    \hfill    ISSN 0418-9833} \\
{\tt December 2008}                  \\
\end{flushleft}

\vspace{2cm}
\begin{center}
\begin{Large}

{\bf A General Search for New Phenomena at HERA \\}

\vspace{2cm}

H1 Collaboration

\end{Large}
\end{center}

\vspace{2cm}

\begin{abstract}

A model--independent search for deviations from the Standard Model prediction is performed using the full $e^\pm p$ data sample collected by the H1 experiment at HERA.
All event topologies involving isolated electrons, photons, muons, neutrinos and jets with transverse momenta above $20$~GeV are investigated in a single analysis. 
Events are assigned to exclusive classes according to their final state. 
A dedicated algorithm is used to search for deviations from the Standard Model in the distributions of the scalar sum of transverse momenta or the invariant mass of final state particles and to quantify their significance. 
Variables related to angular distributions and energy sharing between final state particles are also introduced to study the final state topologies.
No significant deviation from the Standard Model expectation is observed in the phase space covered by this analysis.

\end{abstract}

\vspace{1.5cm}

\begin{center}
Submitted to \PLB
\end{center}

\end{titlepage}

%          THE PAPER DRAFTS HAVE NO AUTHORLIST
%
%          FOR PAPER ISSUED FOR THE FINAL READING 
%          COPY THE AUTHOR AND INSTITUTE LISTS 
%          INTO YOUR AREA
%
% from /h1/iww/ipublications/h1auts.tex 
%          AND UNCOMMENT THE NEXT THREE LINES 
%
\begin{flushleft}

%-- H1AUTS Author list by names 
%-- Status: Mon Dec  1 14:14:10 CET 2008  Number of authors = 255 

F.D.~Aaron$^{5,49}$,           %BUCH-PD        11/06           Aaron               
C.~Alexa$^{5}$,                %BUCH-PD        06/06           Alexa               
V.~Andreev$^{25}$,             %LPI -PD        8/88            Andreev             
B.~Antunovic$^{11}$,           %DESY-LEFT      12/08           Antunovic           
S.~Aplin$^{11}$,               %DESY-LEFT      01/08           Aplin               
A.~Asmone$^{33}$,              %ROME-ST        07/2            Asmone              
A.~Astvatsatourov$^{4}$,       %BRUX-LEFT      01/08           Astvatsatourov      
S.~Backovic$^{30}$,            %PODG-PD        03/2            Backovic            
A.~Baghdasaryan$^{38}$,        %YERE-PD        09/03           Baghdasaryana       
E.~Barrelet$^{29}$
% ...

%-- H1 Institutes 
\bigskip{\it
% ...
$ ^{4}$ Inter-University Institute for High Energies ULB-VUB, Brussels;
          Universiteit Antwerpen, Antwerpen; Belgium$^{ c}$ \\
 $ ^{5}$ National Institute for Physics and Nuclear Engineering (NIPNE) ,
          Bucharest, Romania \\
% ...
$ ^{11}$ DESY, Hamburg, Germany \\
% ...
$ ^{25}$ Lebedev Physical Institute, Moscow, Russia$^{ e}$ \\
% ...
$ ^{29}$ LPNHE, Universit\'{e}s Paris VI and VII, IN2P3-CNRS,
          Paris, France \\
 $ ^{30}$ Faculty of Science, University of Montenegro,
          Podgorica, Montenegro$^{ e}$ \\
% ...
$ ^{33}$ Dipartimento di Fisica Universit\`a di Roma Tre
          and INFN Roma~3, Roma, Italy \\
% ...
$ ^{38}$ Yerevan Physics Institute, Yerevan, Armenia \\
% ...
$ ^{49}$ Also at Faculty of Physics, University of Bucharest,
          Bucharest, Romania \\
% ...
$ ^c$ Supported by FNRS-FWO-Vlaanderen, IISN-IIKW and IWT
      and  by Interuniversity
Attraction Poles Programme,
      Belgian Science Policy \\
% ...
$ ^e$ Supported by the Deutsche Forschungsgemeinschaft \\
% ...
}
\end{flushleft}
%
% Please not that the author list may need re-formatting.

\newpage

%%%%%%%%%%%%%%%%%%%%%%%%%%%%%%%%%%%%%%%%%%%%%%%%%%%%%%%%%%%%
\section{Introduction}
%%%%%%%%%%%%%%%%%%%%%%%%%%%%%%%%%%%%%%%%%%%%%%%%%%%%%%%%%%%%%
At HERA electrons\footnote{
  In this paper the term ``electron'' is used generically to refer to both electrons and positrons, unless otherwise stated.}
and protons collide at a centre--of--mass energy of up to $319$~GeV. 
The collected luminosity of high--energy electron-proton interactions gives access to rare processes with cross sections of the order of $0.1$~pb, providing a testing ground for the Standard Model (SM) complementary to $e^+e^-$ and $p\overline{p}$ scattering. 

A large variety of possible extensions to the SM predicts new phenomena which may appear at high energies.
Searches for new physics often compare the data to the predictions of specific models.
A complementary approach is followed in signature based searches by looking for differences between data and SM expectation in various event topologies.
As an advantage, such model independent analyses do not rely on any a priori definition of expected signatures for exotic phenomena.
Therefore, they address the important question of whether unexpected phenomena may occur through a new pattern, not predicted by existing models.
Following this approach, final states corresponding to rare SM processes such as single $W$ boson or lepton pair production have already been investigated at HERA~\cite{Andreev:2003pm,isolep_H2,Aktas:2003jg,Aktas:2003sz,Aaron:2008jh}. Model independent analyses are also performed at the Tevatron~\cite{Abbott:2000fb,Aaltonen:2007dg}.

The present paper reports on a general analysis of all high transverse momentum ($P_T$) final state configurations involving electrons ($e$), muons ($\mu$), jets ($j$), photons ($\gamma$) or neutrinos ($\nu$) in $e^\pm p$ collisions. 
This analysis searches for deviations from the SM prediction in phase space regions where the SM prediction is reliable.
All final states containing at least two particles\footnote{In this context a high $P_T$ jet is also called particle.} ($e$, $\mu$, $j$, $\gamma$, $\nu$) with  $P_T >$~$20$~GeV in the polar angle\footnote{ 
  The origin of the H1 
  coordinate system is the nominal $ep$ interaction point, with 
  the direction of the proton beam defining the positive 
  $z$--axis (forward region). The transverse momenta are measured 
  in the $xy$ plane. 
  The 
  pseudorapidity $\eta$ is related to the polar 
  angle $\theta$ by $\eta = -\ln \, \tan (\theta/2)$.}
range  $10^\circ < \theta < 140^\circ$ are investigated. 
The present analysis follows the strategy of the previous H1 publication~\cite{Aktas:2004pz}.
Selected events are classified into exclusive event classes according to the number and types of particles detected in the final state (e.g.  \ej, \mujnp, \jjjj). 
In a first step the event yields are compared with the SM expectation.
In a second step kinematical distributions are systematically investigated using
a dedicated algorithm~\cite{Aktas:2004pz} which locates 
the region with the largest deviation of the data from the SM prediction.

The complete $e^\pm p$ data sample collected by the H1 experiment at HERA is used.
The data are recorded at an electron beam energy of $27.6$~GeV and proton beam energies of $820$~GeV or $920$~GeV, corresponding to centre--of--mass energies $\sqrt{s}$ of $301$~GeV or $319$~GeV, respectively.
The total integrated luminosity of the data is $463$~pb$^{-1}$, which represents a factor of four increase with respect to the previously published result~\cite{Aktas:2004pz}.
The data comprise $178$~pb$^{-1}$ recorded in $e^-p$ collisions and $285$~pb$^{-1}$ in $e^+p$ collisions, of which $35$~pb$^{-1}$ were recorded at $\sqrt{s} = 301$~GeV.
While the previous general search was dominated by $e^+p$ collision data, a large data set recorded in $e^-p$ scattering is now also analysed.

%%%%%%%%%%%%%%%%%%%%%%%%%%%%%%%%%%%%%%%%%%%%%%%%%%%%%%%%%%%%%
\section{Standard Model Processes and their Simulation}\label{sec:MC}
%%%%%%%%%%%%%%%%%%%%%%%%%%%%%%%%%%%%%%%%%%%%%%%%%%%%%%%%%%%%%

A precise estimate of all processes relevant at high transverse momentum in $ep$ interactions is needed to ensure a reliable comparison to the SM.
Several Monte Carlo (MC) generators are therefore combined to simulate events in all classes. 

At high transverse momenta the dominant SM processes are photoproduction of two jets and neutral current (NC) deep--inelastic scattering (DIS).
Direct and resolved photoproduction of jets and prompt photon production are simulated using the PYTHIA~\cite{Sjostrand:2000wi} event generator. 
The simulation is based on Born level hard scattering matrix elements with radiative QED corrections. 
The RAPGAP~\cite{Jung:1993gf} event generator, which implements the Born, QCD Compton and boson gluon fusion matrix elements, is used to model NC DIS events. 
QED radiative effects arising from real photon emission from both the incoming and outgoing electrons are simulated using the HERACLES~\cite{Kwiatkowski:1990es} program. 
In RAPGAP and PYTHIA, jet production from higher order QCD radiation is simulated using leading logarithmic parton showers. Hadronisation is modelled with Lund string fragmentation~\cite{Sjostrand:2000wi}.
The leading order MC prediction of photoproduction and NC DIS processes with two or more high transverse momentum jets is scaled by a factor of $1.2$ to account for the incomplete description of higher orders in the MC generators~\cite{Adloff:2002au,Aktas:2004pz}. 
Charged current (CC) DIS events are simulated using the DJANGO~\cite{Schuler:yg} event generator, which includes first order leptonic QED radiative corrections based on HERACLES. The production of two or more jets in DJANGO is accounted for using the colour--dipole--model~\cite{Lonnblad:1992tz}. 
Contributions from elastic and quasi--elastic QED Compton scattering are simulated with the WABGEN~\cite{Berger:kp} generator. 
Contributions arising from the production of single $W$ bosons and multi--lepton events are modelled using the EPVEC~\cite{Baur:1991pp} and GRAPE~\cite{Abe:2000cv} event generators, respectively.

All processes are generated with at least ten times the integrated luminosity of the data sample.
Generated events are passed through the GEANT~\cite{Brun:1987ma} based simulation of the H1 apparatus, which takes into account the running conditions of the different data taking periods, and are reconstructed and analysed using the same program chain as is used for the data.

%%%%%%%%%%%%%%%%%%%%%%%%%%%%%%%%%%%%%%%%%%%%%%%%%%%%%%%%%%%%%
\section{Experimental Conditions}
%%%%%%%%%%%%%%%%%%%%%%%%%%%%%%%%%%%%%%%%%%%%%%%%%%%%%%%%%%%%%

A detailed description of the H1 experiment can be found in~\cite{Abt:h1}.
Only the detector components relevant to the
present analysis are briefly described here.  
The Liquid Argon (LAr) calorimeter~\cite{Andrieu:1993kh} covers the polar angle range
$4^\circ < \theta < 154^\circ$ with full azimuthal acceptance.
Electromagnetic shower energies are measured with a precision of
$\sigma (E)/E \simeq 11\%/ \sqrt{E/\mbox{GeV}} \oplus 1\%$ and hadronic energies
with $\sigma (E)/E \simeq 50\%/\sqrt{E/\mbox{GeV}} \oplus 2\%$, as measured in test beams~\cite{Andrieu:1993tz,Andrieu:1994yn}.
In the backward region, energy measurements are provided by a lead/scintillating--fibre (SpaCal) calorimeter~\cite{Appuhn:1996na} covering the range $155^\circ < \theta < 178^\circ$.
The central ($20^\circ < \theta < 160^\circ$)  and forward ($7^\circ < \theta < 25^\circ$)  inner tracking detectors are used to
measure charged particle trajectories and to reconstruct the interaction
vertex.
The innermost central proportional chamber, CIP~\cite{Muller:1992jk,Becker:2007ms}  ($9^\circ < \theta < 171^\circ$) is used together with tracking detectors to veto charged particles for the identification of photons.
The LAr calorimeter and inner tracking detectors are enclosed in a super--conducting magnetic
coil with a field strength of $1.16$~T.
From the curvature of charged particle trajectories in the magnetic field, the central tracking system provides transverse momentum measurements with a resolution of $\sigma_{P_T}/P_T = 0.005 P_T / \rm{GeV} \oplus 0.015$~\cite{Kleinwort:2006zz}.
The return yoke of the magnetic coil is the outermost part of the detector and is
equipped with streamer tubes forming the central muon detector
($4^\circ < \theta < 171^\circ$).
In the forward region of the detector ($3^\circ < \theta < 17^\circ$) a set of drift chambers detects muons and measures their momenta using an iron toroidal magnet.
The luminosity is determined from the rate of the Bethe--Heitler process $ep \rightarrow ep \gamma$,
measured using a photon detector located close to the beam pipe at $z=-103~{\rm m}$, in the backward direction.

The main trigger for events with high transverse momentum is provided by the LAr calorimeter~\cite{Adloff:2003uh}.
Events with an electromagnetic deposit  (electron or photon) in the LAr with an energy greater than $10$~GeV are detected by the LAr trigger with an efficiency of about $100$\%~\cite{nikiforov}.
Events are also triggered by jets only, with a trigger efficiency above $95$\% for $P_T^{\rm{jet}} > 20$~GeV and nearly $100$\% for $P_T^{\rm{jet}} > 25$~GeV~\cite{matti}.
For events with missing transverse energy of $20$~GeV, the trigger efficiency is about $90$\% and increases above $95$\% for missing transverse energy above $30$~GeV~\cite{trinh}.
The trigger for events with only muons is based on single muon signatures from the central muon detector, combined with signals from the central tracking detector. The trigger efficiency is about $95$\% for di--muon events with muon transverse momenta larger than $15$~GeV~\cite{Aaron:2008jh}.
%

%%%%%%%%%%%%%%%%%%%%%%%%%%%%%%%%%%%%%%%%%%%%%%%%%%%%%%%%%%%%%
\section{Data Analysis}
%%%%%%%%%%%%%%%%%%%%%%%%%%%%%%%%%%%%%%%%%%%%%%%%%%%%%%%%%%%%%

\subsection{Event Reconstruction and Particle Identification}\label{sec:part_def}

In order to remove background events induced by cosmic showers and other non--$ep$ sources, the event vertex is required to be within $35$~cm in $z$ of the nominal interaction point. In addition, topological filters and timing vetoes are applied~\cite{negri}.

Calorimetric energy deposits and tracks are used to look for electron, photon and muon candidates.
Electron and photon candidates are characterised by compact and isolated electromagnetic showers in the LAr calorimeter.  
The identification of muon candidates is based on a track measured in the inner tracking systems associated with signals in the muon detectors~\cite{Andreev:2003pm}.
Calorimeter energy deposits and tracks not previously identified as electron, photon or muon candidates are used to form combined cluster--track objects, from which the hadronic final state is reconstructed~\cite{matti,benji}.
Jet candidates with a minimum transverse momentum of $2.5$~GeV are reconstructed from these combined cluster--track objects using an inclusive $k_T$ algorithm~\cite{Ellis:1993tq,Catani:1993hr} with a $P_T$ weighted recombination scheme in which the jets are treated as massless.
The missing transverse momentum  $P_T^{\rm{miss}}$ of the event is derived from all detected particles and energy deposits in the event.
In events with large $P_T^{\rm{miss}}$, a neutrino candidate is reconstructed.  
The four--vector of this neutrino candidate is calculated assuming  transverse momentum conservation and the relation $\sum_i (E^i - P_{z}^{i}) + (E^\nu - P_{z}^{\nu}) = 2 E^0_e = 55.2$~GeV, where the sum runs over all detected particles, $P_{z}$ is the momentum along the proton beam axis and $E^0_e$ is the electron beam energy.
The latter relation holds if no significant losses are present in the electron beam direction.

Additional requirements are applied to ensure an unambiguous identification of particles, while retaining good efficiencies. 
Strict isolation criteria are applied in order to achieve high purities in all event classes.

For electrons, the calorimetric energy measured within a distance in the pseudorapidity--azimuth $(\eta, \phi)$ plane $R=\sqrt{\Delta \eta^2 + \Delta \phi^2} < 0.75$ around the candidate is required to be below $2.5$\% of its energy.
In the region of angular overlap between the LAr and
the central tracking detectors ($20^\circ < \theta < 140^\circ$), hereafter referred to as the central region, the calorimetric electron identification is complemented by tracking information.  
In this region it is required that a well measured track
geometrically matches the centre--of--gravity of the electromagnetic cluster within a distance of closest approach (DCA) of $12$~cm.  
Furthermore, the distance from the first measured track point in the central drift
chambers to the beam axis is required to be below $30$~cm in order to
reject photons that convert late in the central tracker material. 
In the central region, the transverse momentum of the associated electron track $P_T^{e_{tk}}$ is required to match the
calorimetric measurement $P_T^e$ such that $1/P_T^{e_{tk}} - 1/P_T^e < 0.02$~GeV$^{-1}$ in order to reject hadronic showers.  
In the forward region ($10^\circ < \theta < 20^\circ$), a wider calorimetric isolation cone of $R < 1$ is required to reduce the contribution of fake electrons from hadrons.
In this forward region, at least one track is required to be present with a DCA $< 12$~cm. 
The presence of at least one hit in the CIP, associated to the electron trajectory, is also required.
Finally, the electron is required to be isolated from any other well measured track by a distance $R > 0.5$ ($R > 1$) to the electron direction in the central (forward) region.
The resulting electron identification efficiency is $\sim 80$\% in the central region and $\sim 40$\% in the forward region, determined from NC DIS events.

The identification of photons relies on the same calorimetric isolation criteria as used in the electron identification.
Vetoes on any track pointing to the electromagnetic cluster 
are applied. 
No track with a DCA to the cluster below $24$~cm or within $R < 0.5$ should be present.
An additional veto on any hits in the CIP associated to the electromagnetic cluster is applied.
Furthermore, each photon must be isolated from jets by $R > 0.5$.
The resulting photon identification efficiency as derived using 
elastic QED Compton events is $\sim 95\%$ in the central region and $\sim 50$\% in the forward region.

A muon should have no more than $5$~GeV deposited in a
cylinder, centred on the muon track direction, of radius $25$~cm and $50$~cm in the electromagnetic and hadronic sections of the LAr calorimeter, respectively.
Misidentified hadrons are strongly suppressed by requiring that the muon be separated from the closest jet and from any further track by $R > 1$.
In di--muon events, the opening angle between the two muons is required to be smaller than $165^\circ$, in order to remove muons originating from cosmic rays.
The efficiency to identify muons is $\sim 90$\%~\cite{Aaron:2008jh}.

The scattered electron may be misidentified as a hadron and reconstructed as a jet.
To reject fake jet candidates, the first radial moment of the jet transverse energy~\cite{Giele:1997hd,frising} is required to be greater than $0.02$ and the quantity
$M^{\rm{jet}}/P_T^{\rm{jet}}$ greater than $0.1$~\cite{Adloff:2002au,frising}, where the invariant mass $M^{\rm{jet}}$ is obtained using the four--vector sum of all particles belonging to the jet. 
If the fraction of the jet energy contained in the 
electromagnetic part of the LAr calorimeter is greater than $0.9$, the above
criteria are tightened to $0.04$ and $0.15$, respectively. 
%
%The jet selection efficiency is $\sim 97\%$~\cite{frising}.
These requirements are fulfilled  by $\sim 97\%$ of the jets~\cite{frising}.

Missing transverse momentum, which is the main signature for neutrinos, may arise from mis--measurement of particles. 
By requiring $\sum_i \left(E^i-P_{z}^{i}\right)<48$~GeV, fake neutrino 
candidates from NC DIS processes are rejected.
If exactly one electron or muon candidate is found, a neutrino is only assigned to an event if $\Delta \phi_{(l-X_{h})}<160^\circ$, where $\Delta \phi_{(l-X_{h})}$ is the difference in azimuthal angle between the lepton $l$ and the direction of the hadronic final state $X_{h}$.

\subsection{Event Selection and Classification}

The common phase space for electrons, photons, muons and jets is defined 
by $10^\circ<\theta<140^\circ$ and $P_T > 20$~GeV. 
The  neutrino phase space is defined as missing transverse momentum above $20$~GeV and  
$\sum_i \left(E_i-P_{z,i}\right)<48$~GeV. 
All particles with $P_T > 20$~GeV, including the neutrino defined by its reconstructed four--vector, are required to be isolated with respect to each other by a minimum distance $R>1$. 
The particles satisfying these requirements are referred to as bodies.
The events are sorted depending on the number and types of bodies into exclusive event classes. 
All possible event classes with at least two bodies are investigated. 
Only the \munp~event class is discarded from the analysis. This class is dominated by events in which a poorly reconstructed muon gives rise to missing transverse momentum, which fakes the neutrino signature. 

Based on these identification criteria, purities have been 
derived for each event class. 
Purity is defined as the ratio of SM events reconstructed 
in the event class in which they are generated to the total number of 
reconstructed events in this class. 
Most purities are found to be above $60\%$ and are close to $100\%$ for the 
\jj, \ej, \jnp~and \mumu~event classes.

\subsection{Systematic Uncertainties}
\label{sec:syst}

The following experimental systematic uncertainties are considered:

\begin{itemize}
\item The uncertainty on the electromagnetic energy scale varies depending on the polar angle from $0.7$\% in the central region to $2$\% in the forward region. The polar angle measurement  uncertainty of electromagnetic clusters is $3$~mrad. 
The identification efficiency of electrons (photons) is known with an uncertainty of $3$\% ($5$\%) to $5$\% ($10$\%), depending on the polar angle. 
\item The scale uncertainty on the transverse momentum of high $P_T$ muons is $2.5$\%~\cite{Aaron:2008jh}. The uncertainty on the reconstruction of the muon polar angle is $3$~mrad. The identification efficiency of muons is known with an uncertainty of $5$\%.
\item The jet energy scale is known within $2$\%~\cite{trinh}. The uncertainty on the jet polar angle determination is $10$ mrad.
\item The uncertainty on the trigger efficiency is estimated to be $6$\% if only muons are present in the final state and $3$\% in all other cases.
\item The luminosity measurement has an uncertainty of $3$\%.
\end{itemize}

The effects of the above uncertainties on the SM expectation are determined by varying the experimental quantities by $\pm 1$ standard deviation in the MC samples and propagating these variations through the whole analysis chain.

Additional model uncertainties are attributed to the SM Monte Carlo generators described in section~$\ref{sec:MC}$.
An error of $10$\% is attributed to NC and CC DIS processes with only one high $P_T$ jet.
To account for the uncertainty on higher order QCD corrections, an error of $15$\% on the normalisation of NC DIS and photoproduction processes with at least two high $P_T$ jets is considered. 
The normalisation uncertainty of CC DIS processes with at least two high $P_T$ jets is estimated to be $20$\%~\cite{trinh}.
For each additional jet produced by parton shower processes, a further theoretical error of $20$\% is added~\cite{wessels}, for example $20$\% for the \jjj~ event class.

The error on the elastic and quasi--elastic QED Compton cross sections is conservatively estimated to be $5$\%. 
The error on the inelastic QED Compton cross section is $10$\%.
The errors attributed to lepton--pair and $W$ production are $3$\% and $15$\%, respectively.
An uncertainty of $30$\% on the simulation of radiative CC DIS events is considered to account for the lack of QED radiation from the quark line in the DJANGO generator. This uncertainty is estimated for the specific phase space of the analysis by a comparison of the DJANGO result to the calculated cross section of the $e^- p {\rightarrow} \nu_e \gamma X$ process~\cite{Helbig:1991iw}.
An uncertainty of $50\%$ is added to the prediction for NC DIS events with measured missing transverse momentum above $20$~GeV and a high $P_T$ electron. 
This uncertainty is estimated by a comparison of the missing transverse momentum distribution of data events containing a low $P_T$ electron ($P_T^e < 20$~GeV) with the SM prediction~\cite{wessels}.

The total error on the SM prediction is determined by adding the effect of all model and experimental systematic uncertainties in quadrature.

%%%%%%%%%%%%%%%%%%%%%%%%%%%%%%%%%%%%%%%%%%%%%%%%%%%%%%%%%%%%%%%%%%%%%%%%%
\section{Results}
%%%%%%%%%%%%%%%%%%%%%%%%%%%%%%%%%%%%%%%%%%%%%%%%%%%%%%%%%%%%%%%%%%%%%%%%%

\subsection{Event Yields}

\begin{figure}[p]
  \center
  \includegraphics[width=0.99\textwidth,angle=-90]{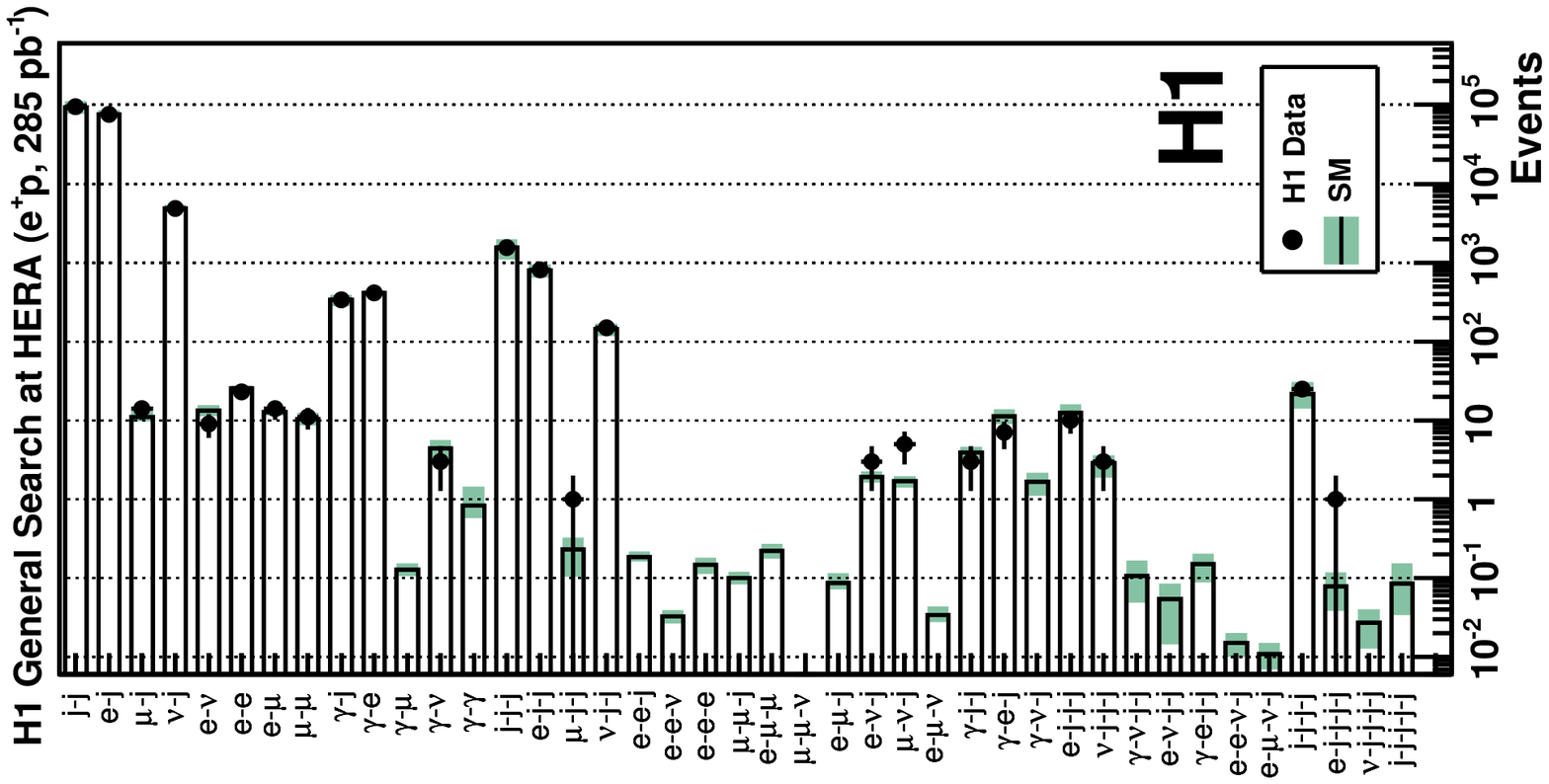}\put(-9.7,-110){{(a)}}
  \includegraphics[width=0.99\textwidth,angle=-90]{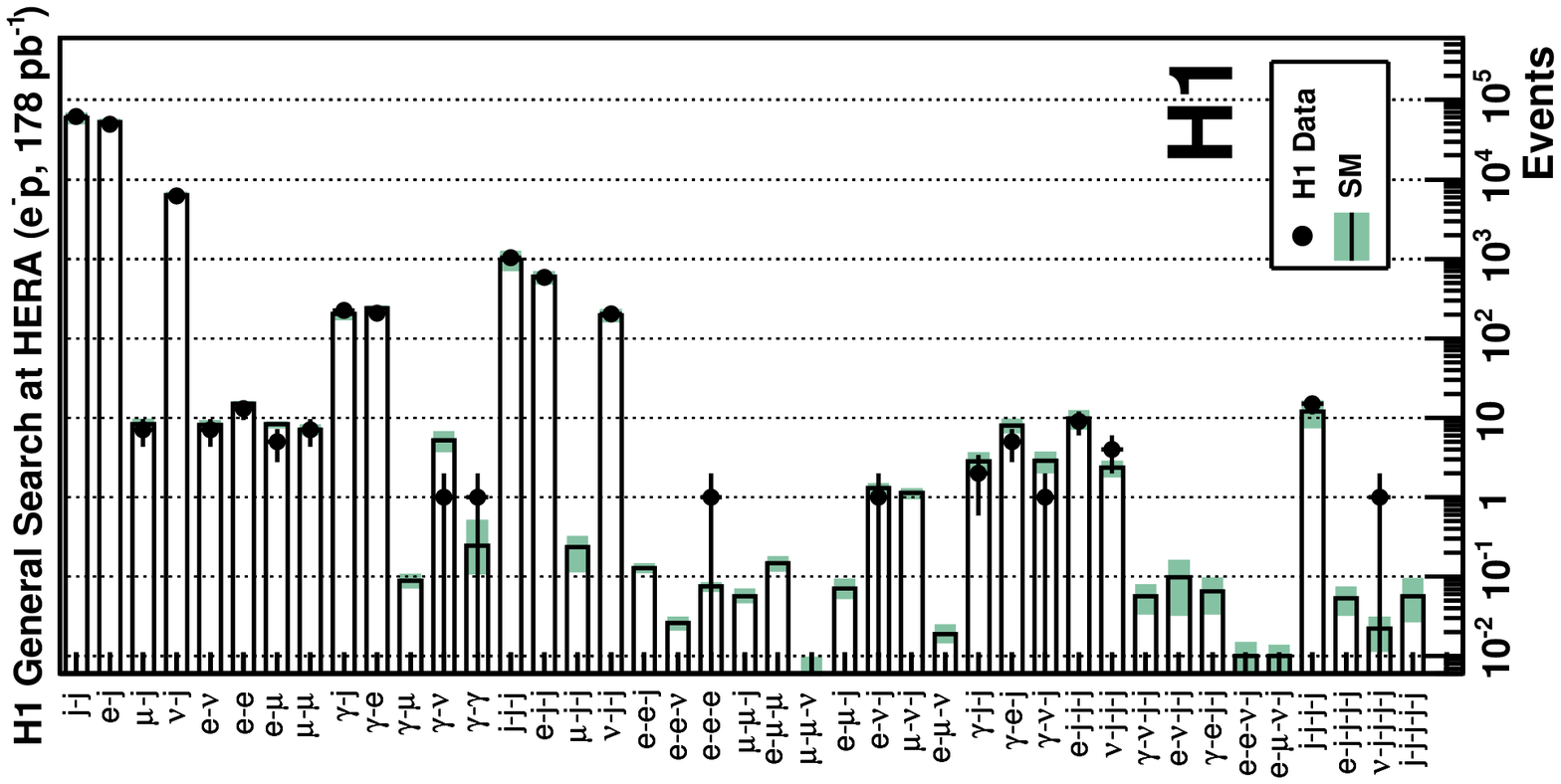}\put(-9.7,-110){{(b)}}
  \caption{The data and the SM expectation for all event classes 
    with observed data events or a SM expectation greater than $0.01$ events for $e^+p$ collisions (a) and $e^-p$ collisions (b). 
    The error bands on the predictions include model uncertainties and 
    experimental systematic errors added in quadrature.
  }
  \label{fig:summaryplot}
\end{figure}

The event yields for all event classes are presented for the data and SM expectation in figures~\ref{fig:summaryplot}(a) and \ref{fig:summaryplot}(b) for $e^+p$ and $e^-p$ collisions, respectively. 
All event classes with observed data events or with a SM expectation greater than $0.01$ events are shown. 
The corresponding observed and predicted event yields for all $e^\pm p$ data are summarised in table~\ref{tab:yields}.
Events are observed in $27$ classes and a good description of the number of observed data events by the SM prediction is seen in each class.

The \jj, \jjj~and \jjjj~event classes are dominated by photoproduction processes.
No event with five jets is observed.
The SM prediction of the \ej, \ejj, \ejjj~and \ejjjj~event classes is dominated by NC DIS processes.
One event, already discussed in a previous H1 publication~\cite{Aktas:2004pz}, is observed in the \ejjjj~event class and compares to a SM prediction of $0.13 \pm 0.06$.
The \jnp, \jjnp, \jjjnp~and \jjjjnp~event classes mainly contain events from CC DIS processes.
One event is observed in the \jjjjnp~event class compared to a SM expectation of $0.05 \pm 0.02$.

Events from QED Compton processes populate the \epho~event class as well as the \ejpho~event class in the case of inelastic events.
The \jpho~event class corresponds to prompt photon events. The purity in this class is moderate ($\sim 50$\%) due to the high background from misidentified electrons in NC DIS.
A slight deficit of data events is observed in the radiative CC DIS classes \nppho~and \jnppho.

Lepton pair production from $\gamma\gamma$ processes dominates in event classes with several leptons (\ee, \mumu, \emu~and \eee).
Compared to the results of a previous study of multi--lepton topologies~\cite{Aaron:2008jh}, the phase space of the present analysis is restricted to higher $P_T$ and extended to forward polar angles down to $10^\circ$.
All multi--lepton events mentioned in~\cite{Aaron:2008jh} and located in the phase space of this analysis are found.
The \ee~event class contains $7$ events with an invariant mass $M_{ee} > 100$~GeV compared to a SM expectation of $3.4 \pm 0.5$ of which $69$\% are from lepton pair processes.
The \eee~event class contains one event compared to a SM expectation of $0.22 \pm 0.04$.

The prediction for the event classes \mujnp~and \ejnp~consists mainly of high $P_T$ single $W$ production with subsequent leptonic decay.
In the \mujnp~(\ejnp) event class $5$ ($4$) events are observed, with a SM expectation of $2.8 \pm 0.5$ ($3.2 \pm 0.5$).
Two events classified as \mujnp~in the previous analysis~\cite{Aktas:2004pz} now migrate to \muj~and \jnp~event classes, respectively, due to improvements in the energy and momentum reconstruction.
Events arising from $W$ production also enter in the \enp~event class.
In this class $16$ events are observed compared to an expectation of $21.5 \pm 3.5$, of which about $90$\% is due to $W$ production processes.

\subsection{Event Topology}\label{sec:topo}

\begin{figure}[p]
  \includegraphics[width=\textwidth]{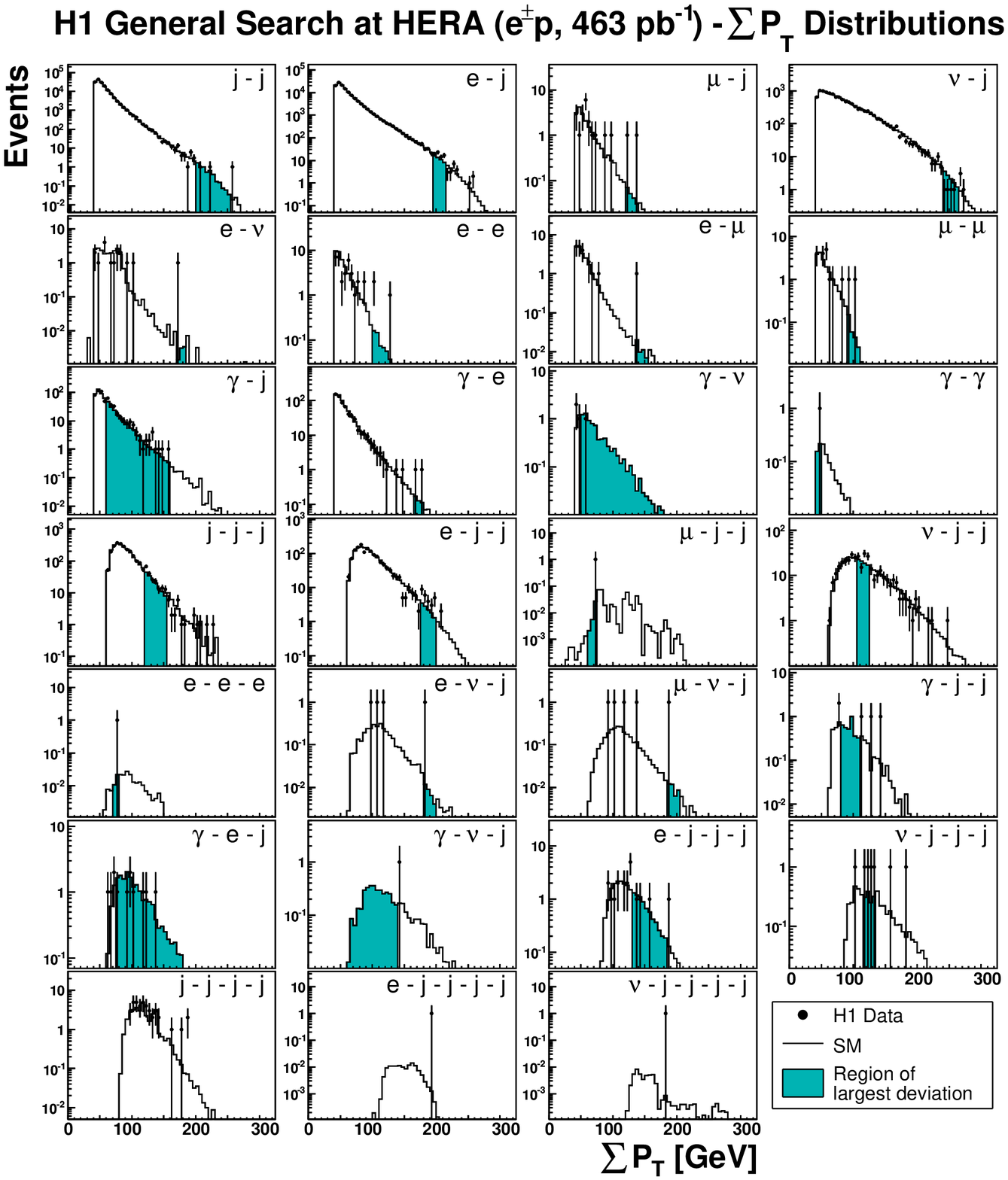}
  \caption{The number of data events and the SM expectation as a 
    function of $\SPT$ for classes with at least one event. 
    The shaded areas show the regions of largest deviation 
    identified by the search algorithm. No such search is performed for the \jjjj, \ejjjj~and \jjjjnp~classes.}
  \label{fig:et_dist}
\end{figure}

\begin{figure}[p]
  \includegraphics[width=\textwidth]{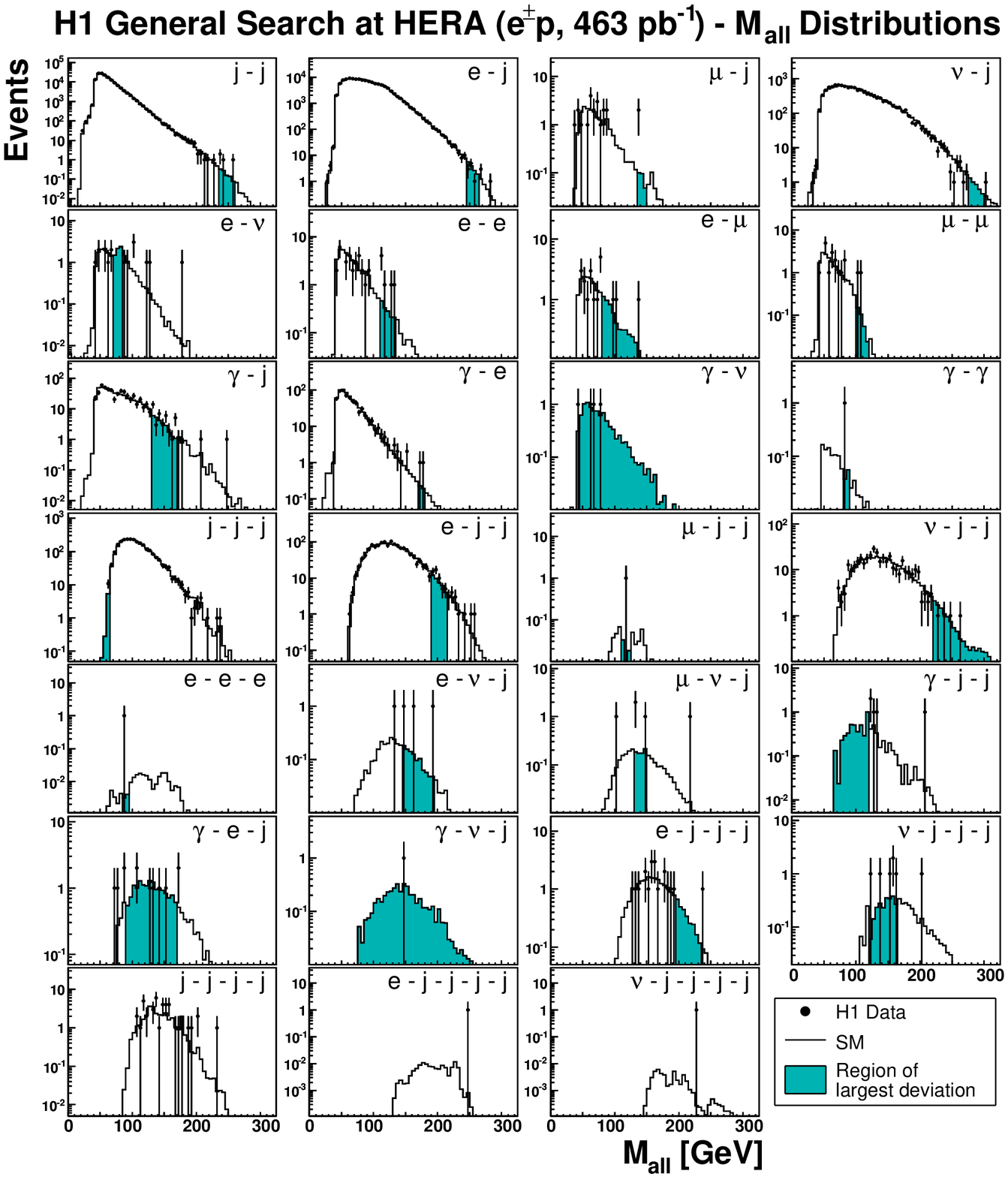}
  \caption{The number of data events and the SM expectation as a
    function of $\Mall$ for classes with at least one event. 
    The shaded areas show the regions of largest deviation 
    identified by the search algorithm. No such search is performed for the \jjjj, \ejjjj~and \jjjjnp~classes.}
  \label{fig:mall_dist}
\end{figure}

The distributions of the scalar sum of transverse momenta $\SPT$ and of the invariant mass $\Mall$ of all bodies are presented in figures~\ref{fig:et_dist} and \ref{fig:mall_dist}, respectively, for classes with at least one event.
The data are in agreement with the SM prediction.
In particular, multiple jets topologies, which are sensitive to QCD radiation, are well described by the simulation.

\begin{figure}[htbp] 
%   \begin{center}
\includegraphics[width=.5\textwidth]{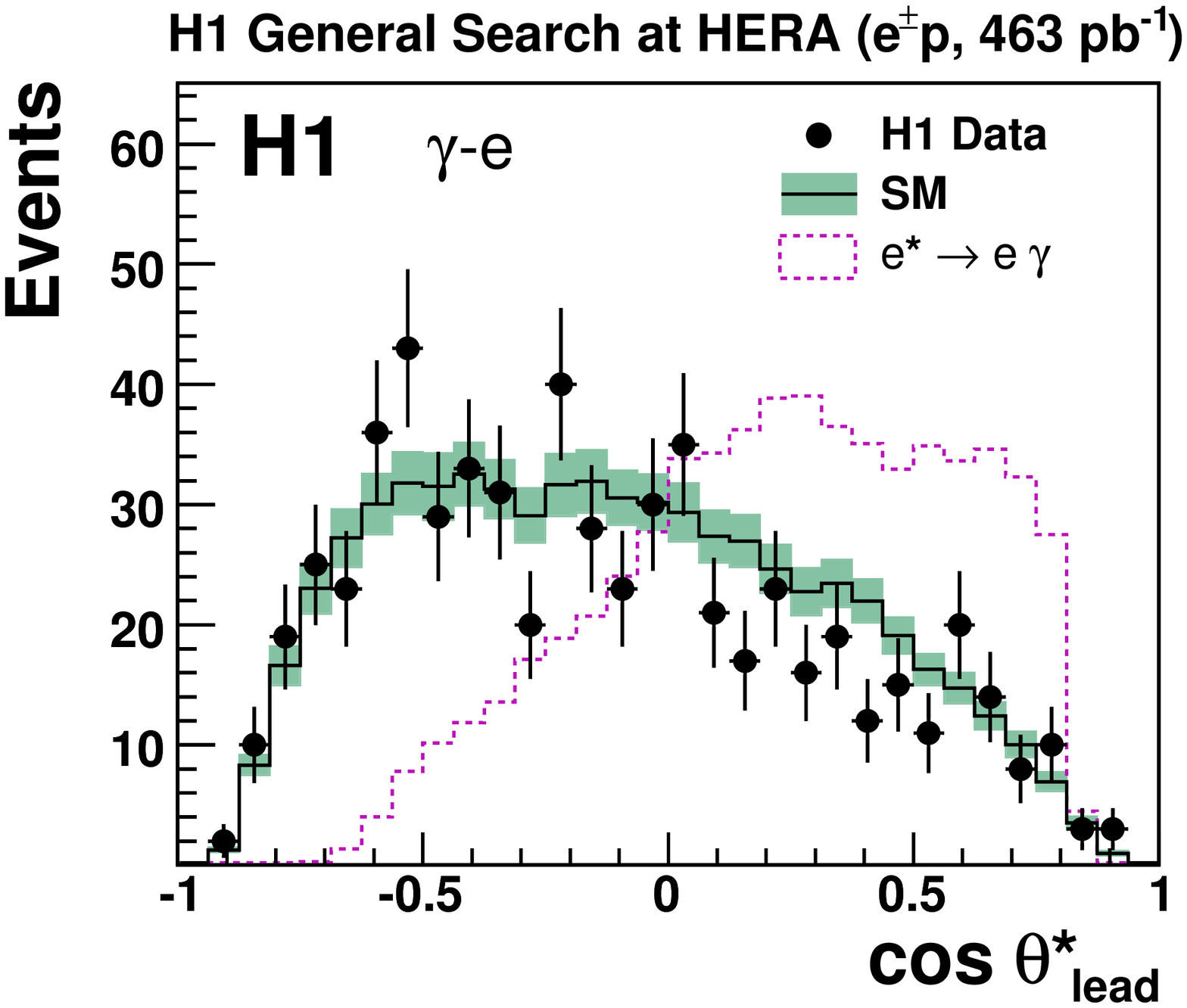}\put(-12,42) {{(a)}}
\includegraphics[width=.5\textwidth]{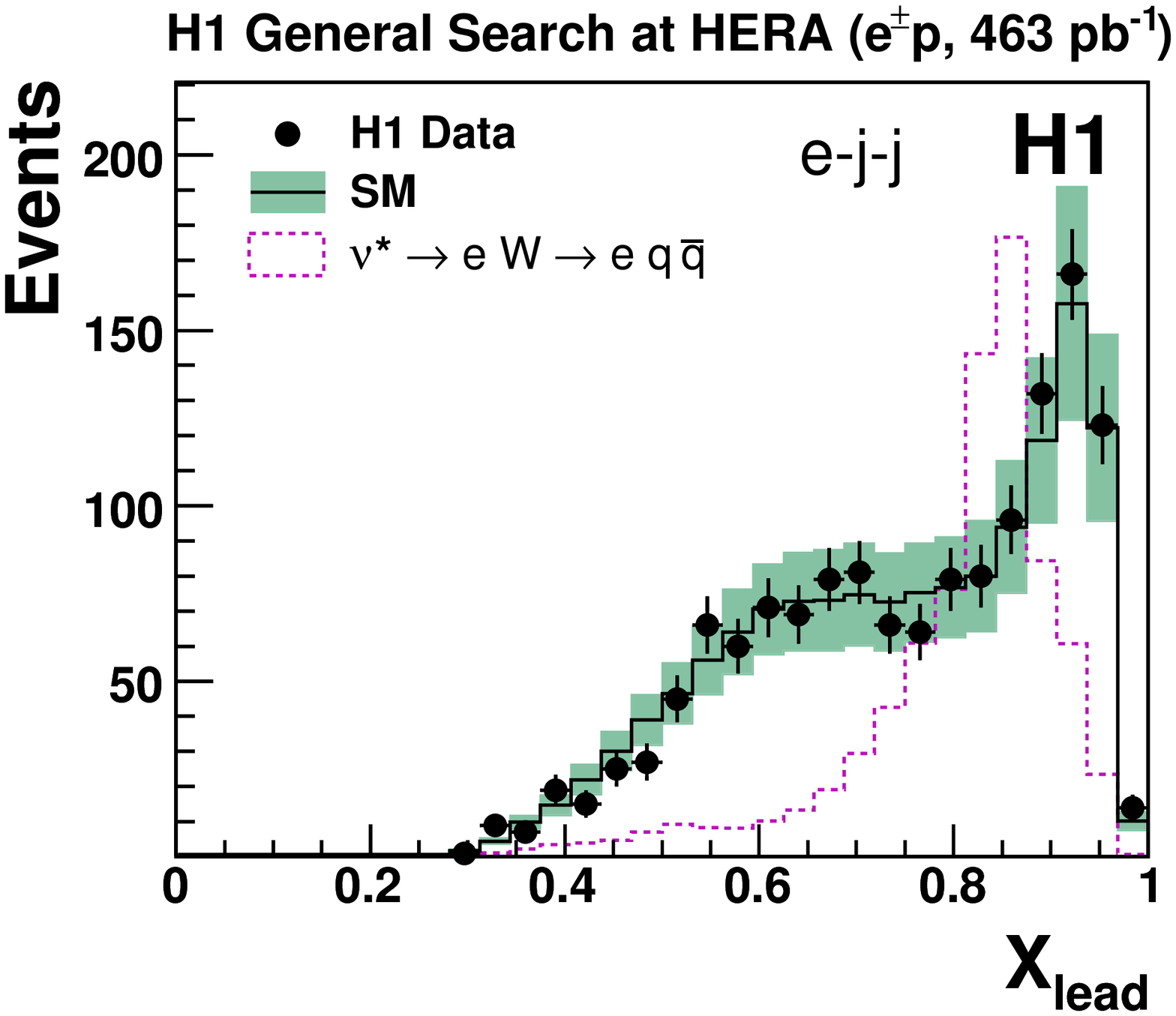}\put(-64,42) {{(b)}}\\
%\end{center}      
 \caption{The $\CosTh$ distribution in the \epho~event class (a) and the $\Xlead$ distribution in the \ejj~event class (b). The points correspond to the observed data events and the histograms to the SM expectation. The error bands on the SM prediction include model uncertainties and experimental systematic errors added in quadrature.
The dashed line represents, with an arbitrary normalisation, the distribution corresponding to an exotic resonance with a mass of  $200$~GeV ($e^*$~\protect{\cite{Collaboration:2008cy}} in (a) and $\nu^*$~\protect{\cite{Aaron:2008xe}} in (b)). 
}
 \label{fig:Topo_ex}  
 \end{figure}

The final state topologies are also evaluated in terms of angular distributions and energy ratios, which are sensitive to spin and decay properties of hypothetical high mass particles. 
Variables used to study the decomposition of the final states, inspired by topological analyses of multi--jet events~\cite{Geer:1995mp}, are defined in the following.
In each event a leading body is selected according to the following priority list between bodies of different types: $\gamma$, $e$, $\mu$, $\nu$, $j$. 
This order of preference allows a better separation of SM background from events originating from a new resonance decaying to a photon or a lepton.
If two bodies of the same type are present, the one with the highest transverse momentum $P_T^*$, relative to the incident proton in the centre--of--mass frame defined by all bodies, is selected.
For classes with exactly two bodies of the same type, the leading body is taken as the one with the highest $P_T$ in the laboratory frame.
The variable $\CosTh$ is then defined as the cosine of the polar angle of the leading body relative to the incident proton in the centre--of--mass frame defined by all bodies.
The variable $\Xlead$ is the energy fraction of the leading body and is defined for systems with three or more bodies as

\beq
\Xlead = \frac{2 E_{\rm{lead}}^*}{\sum_i E_i^*},
\eeq

\noindent where the sum runs over all bodies energies, and $E_{\rm{lead}}^*$ and $E_i^*$ are calculated in the centre--of--mass frame of all bodies.
For events with two bodies, the $\CosTh$ distribution is related to the underlying $2 \rightarrow 2$ matrix element. Therefore, the angular distribution of a particle coming from the decay of a new resonance may be markedly different from that of particles produced in SM processes (see for example \cite{Aktas:2005pr}).
For final states with more than two bodies, $\Xlead$ is a Dalitz variable and related to the dynamics of a possible multi--body decay of a new particle.
The sensitivity of these two variables $\CosTh$ and $\Xlead$ to new physics is tested using different MC samples of exotic processes, for example leptoquarks, excited fermions, or anomalous top production. It has been verified that SM and exotic events exhibit different spectra in these two variables, two examples of which are given in figure~\ref{fig:Topo_ex}.

\begin{figure}[p]
  \center
  \includegraphics[width=\textwidth]{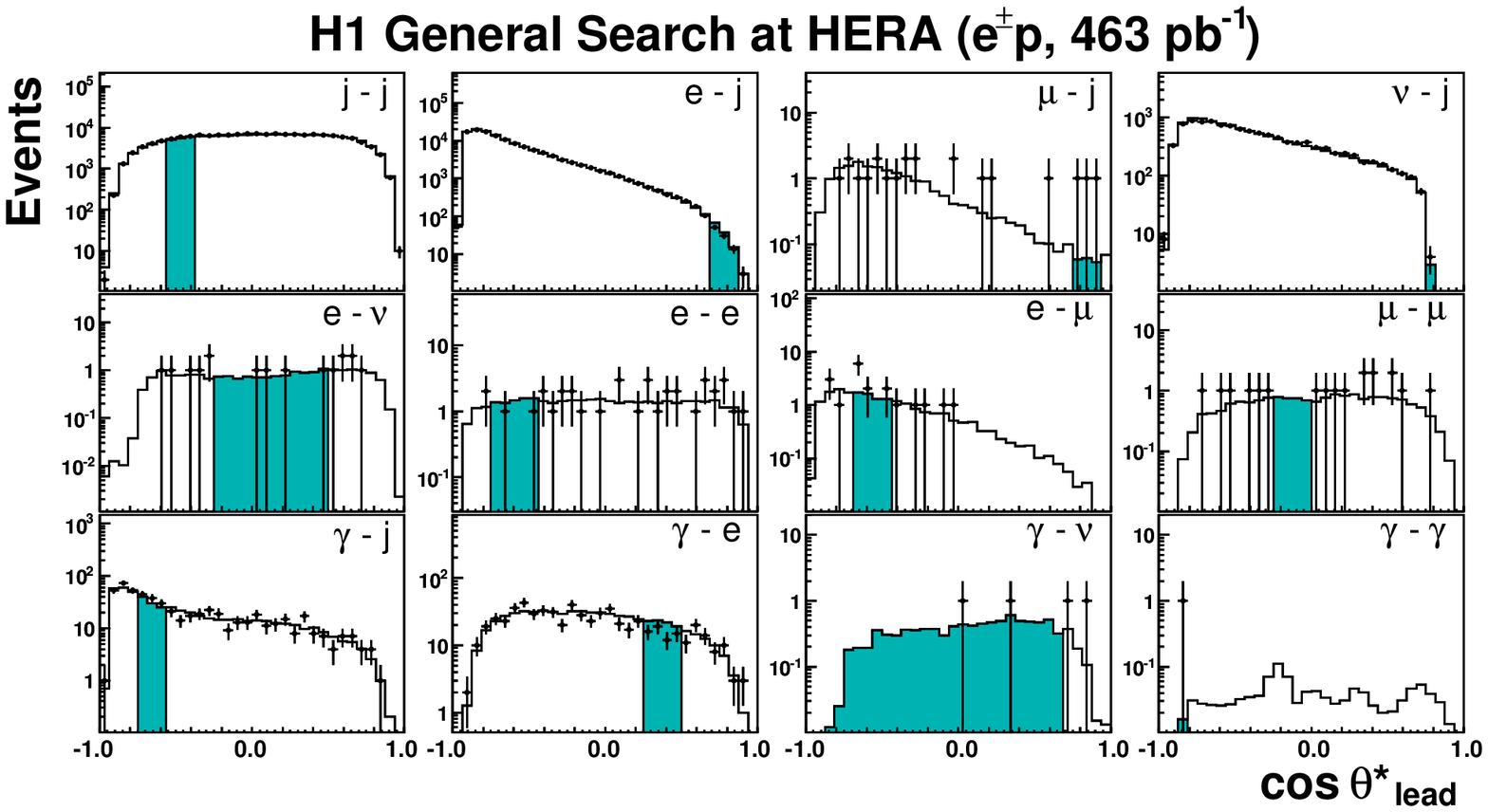}\\
  \includegraphics[width=\textwidth]{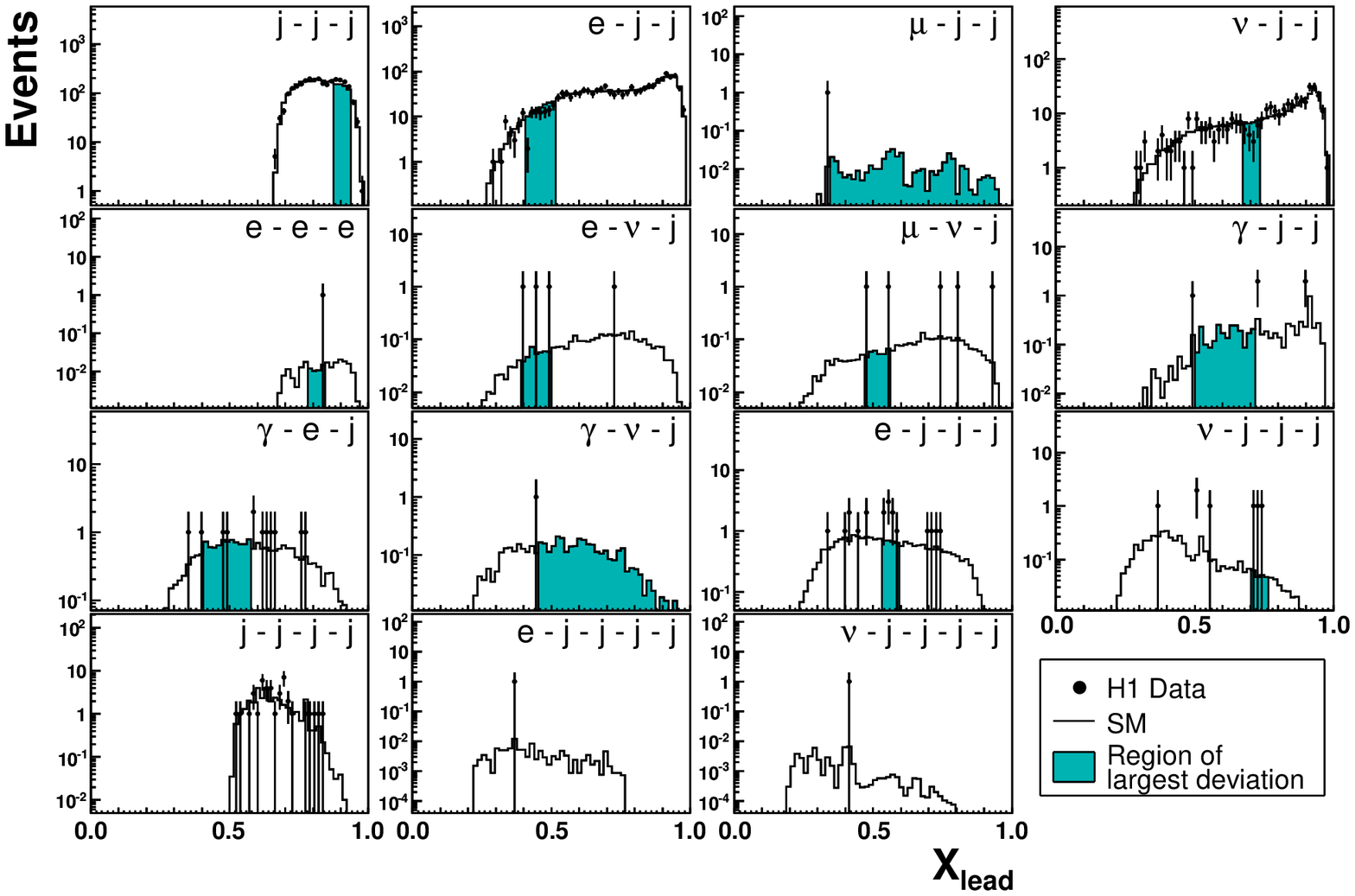}
  \caption{The distribution of $\CosTh$ for event classes with two bodies (top) and of $X_{\rm{lead}}$ for event classes with more than two bodies (bottom). The points correspond to the observed data events and the open histograms to the SM expectation. Only event classes with at least one data event are presented. The shaded areas show the regions of largest deviation identified by the search algorithm. No such search is performed for the \jjjj, \ejjjj~and \jjjjnp~classes.}
  \label{fig:topo_dist}
\end{figure}

The distributions of $\CosTh$ and $\Xlead$ are presented in figure~\ref{fig:topo_dist} for event classes with only two bodies and for event classes with more than two bodies, respectively.
A good overall agreement with the SM prediction is observed in all cases.
This illustrates that the event topology and kinematics, as well as the global variables $\SPT$ and $\Mall$, are well described by the SM.

\subsection{Search for Deviations from the Standard Model}\label{sec:stat}

In order to quantify the level of agreement between the data and the SM expectation and to identify regions of deviations in the $\SPT$, $\Mall$, $\CosTh$ and $\Xlead$ distributions, the search algorithm developed in~\cite{Aktas:2004pz} is used.
A region is defined as a set of connected histogram bins with at least twice the size of the resolution.
A statistical estimator $p$ is defined in order to judge which region is of largest interest. 
This estimator is derived from the convolution of the Poisson probability density function (pdf) to account for statistical errors and a Gaussian pdf to include the effect of systematic uncertainties~\cite{Aktas:2004pz}. 
The value of $p$ gives an estimate of the probability of a fluctuation of the SM expectation upwards (downwards) to at least (at most) the observed number of data events in the region considered.
The region of greatest deviation is the region having the smallest $p$--value, $\pmin$. 
The regions selected by the algorithm in $\SPT$ and $\Mall$ distributions of each class are presented for all $e^\pm p$ data in figures~\ref{fig:et_dist} and~\ref{fig:mall_dist}, respectively.
The corresponding selected regions for $\CosTh$ and $\Xlead$ distributions are shown in figure~\ref{fig:topo_dist}.

The fact that the deviation could have occurred at any point in the distribution is taken into account by calculating the probability $\hat{P}$ to observe a deviation with a $p$--value $\pmin$ at any position in the distribution.
$\hat{P}$ is a measure of the statistical significance of the deviation observed in the data. 
The event class of most interest in the search for anomalies is the one with the smallest $\hat{P}$ value.
Values of $\hat{P}$ larger than $0.01$ indicate event classes where no significant discrepancy between data and the SM expectation is observed. 
The $\hat{P}$ values measured in each of the event classes are listed in table~\ref{tab:yields}.
Due to the uncertainties of the SM prediction in the \jjjj, \ejjjj~and \jjjjnp~event classes, no reliable $\hat{P}$ values can be calculated for them~\cite{Aktas:2004pz} and they are therefore not considered in the search for deviations from the SM.

The overall agreement with the SM can further be quantified by taking into account the large number of event classes in this analysis.
Among all studied classes there is some chance that small $\hat{P}$ values occur. 
This probability can be calculated on a statistical basis with MC experiments. 
A MC experiment is defined as a set of hypothetical data histograms following the SM expectation with an integrated luminosity equal to the amount of data recorded.
The complete search algorithm and statistical analysis are applied to MC experiments analogously as to the data. 
The expectation for the $\hat{P}$ values observed in the data is then given by the distribution of $\hat{P}$ values obtained from all MC experiments.
%

\begin{figure}[p]
  \center
  \includegraphics[width=0.49\textwidth]{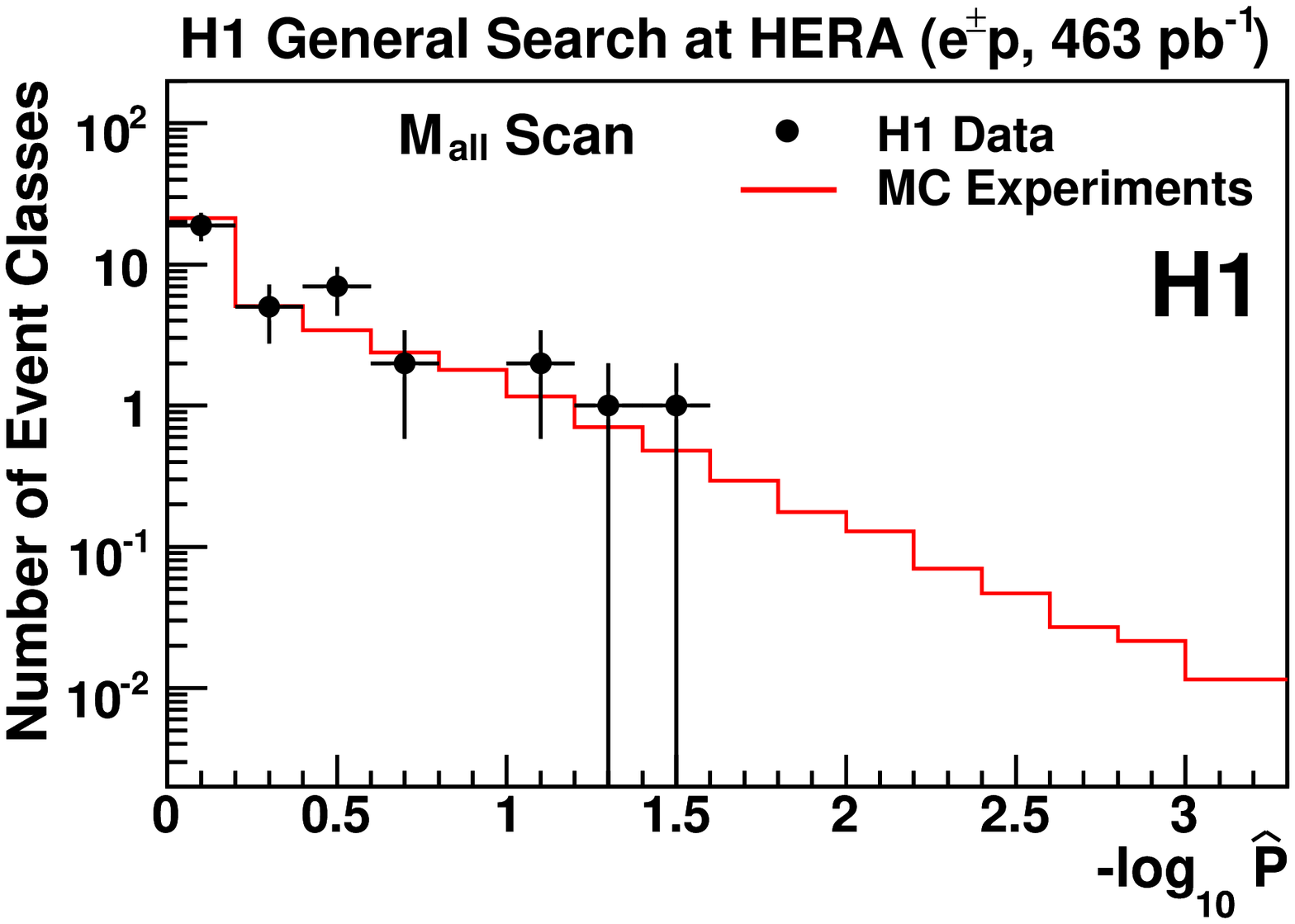}\put(-9.5,30){{(a)}}
  \includegraphics[width=0.49\textwidth]{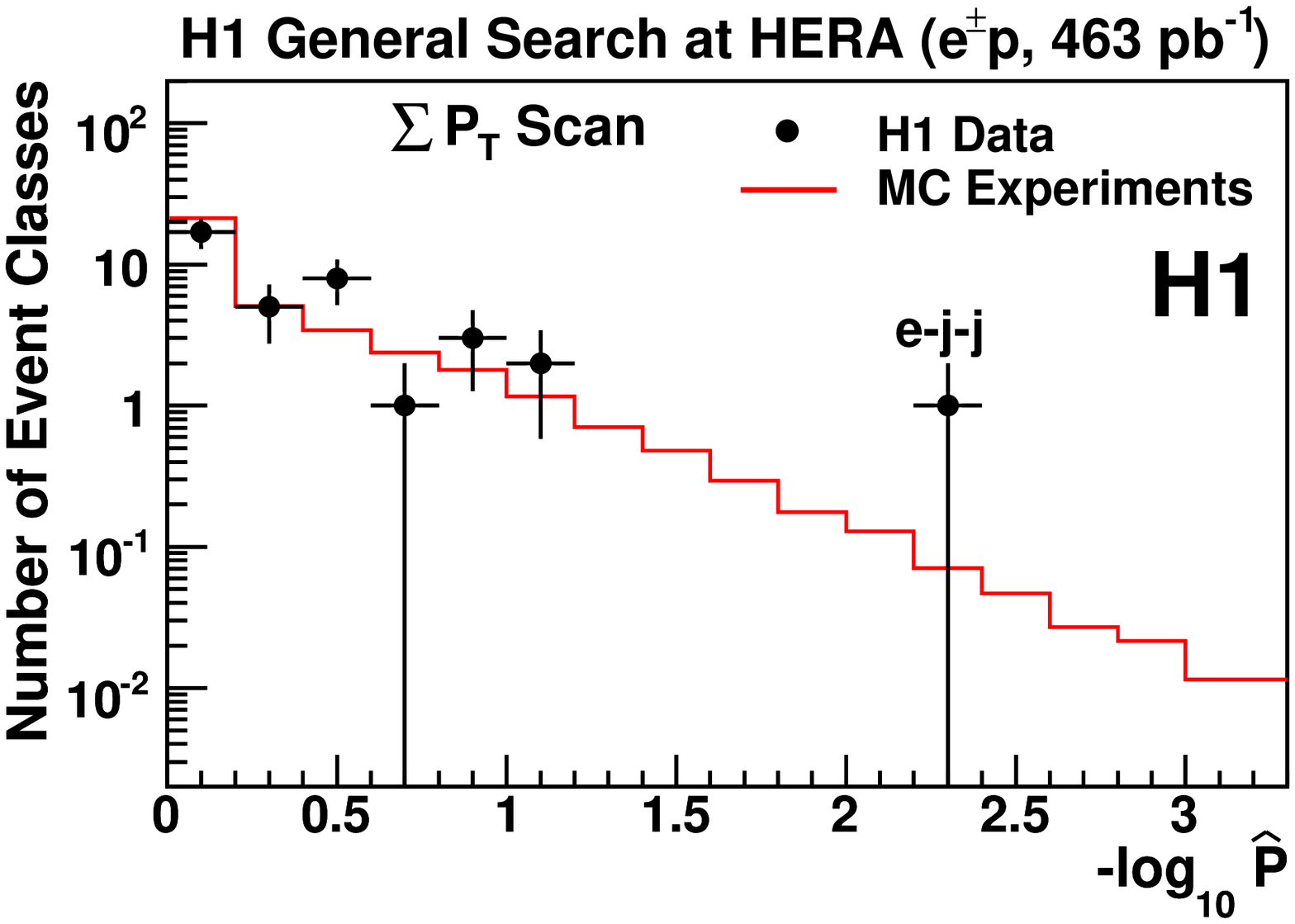}\put(-9.5,30){{(b)}}\\
  \includegraphics[width=0.49\textwidth]{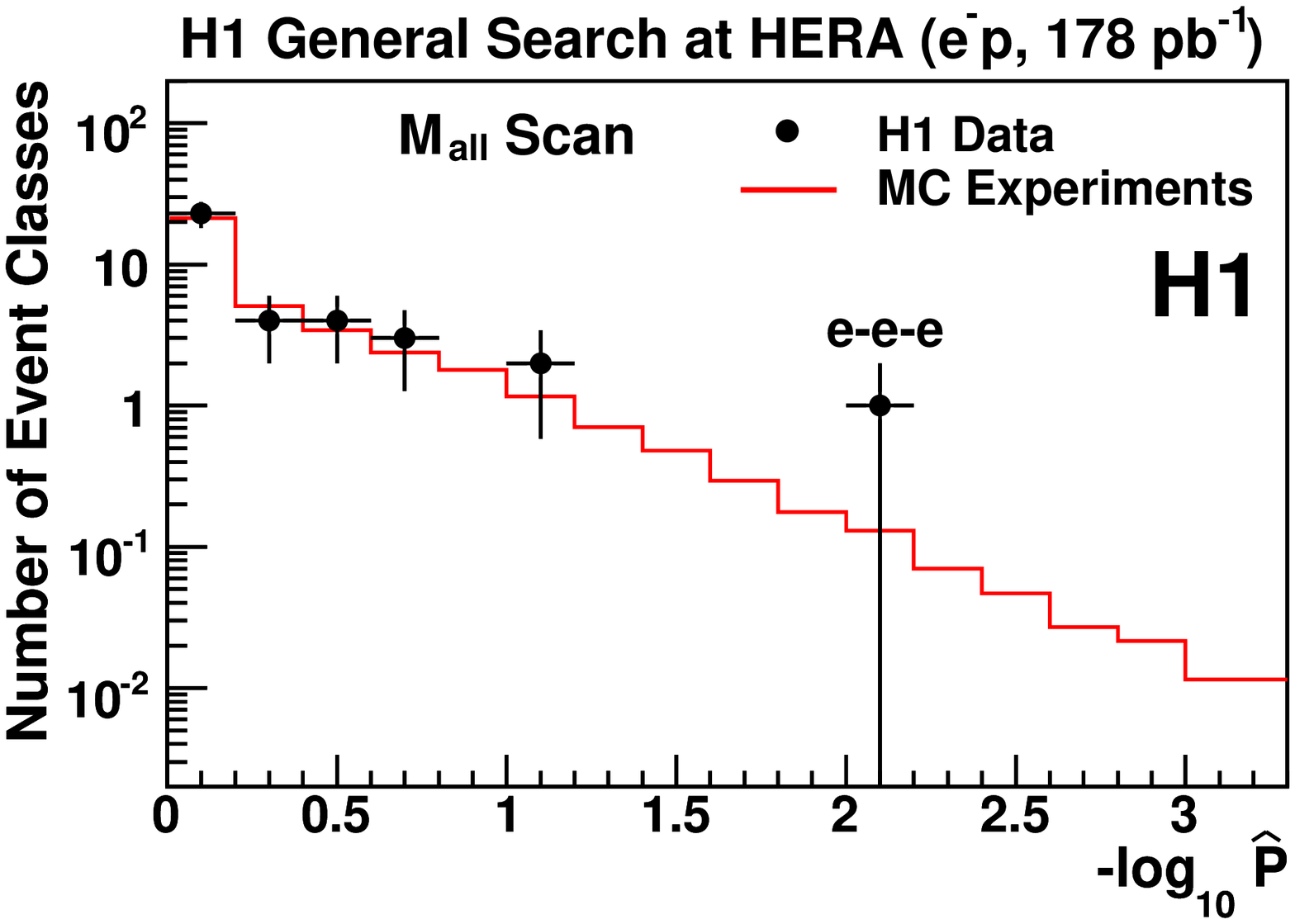}\put(-9.5,30){{(c)}}
  \includegraphics[width=0.49\textwidth]{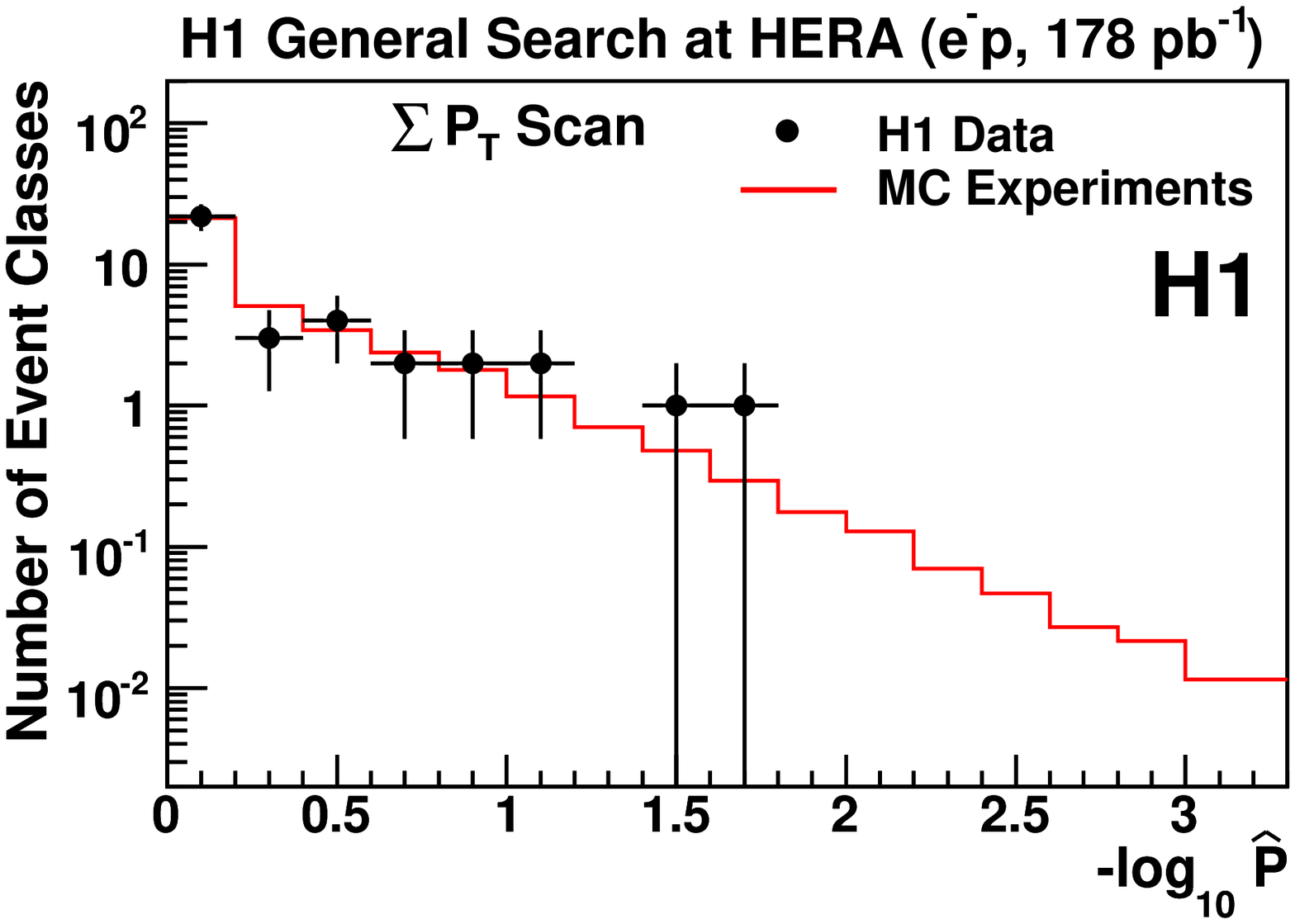}\put(-9.5,30){{(d)}}\\  
  \includegraphics[width=0.49\textwidth]{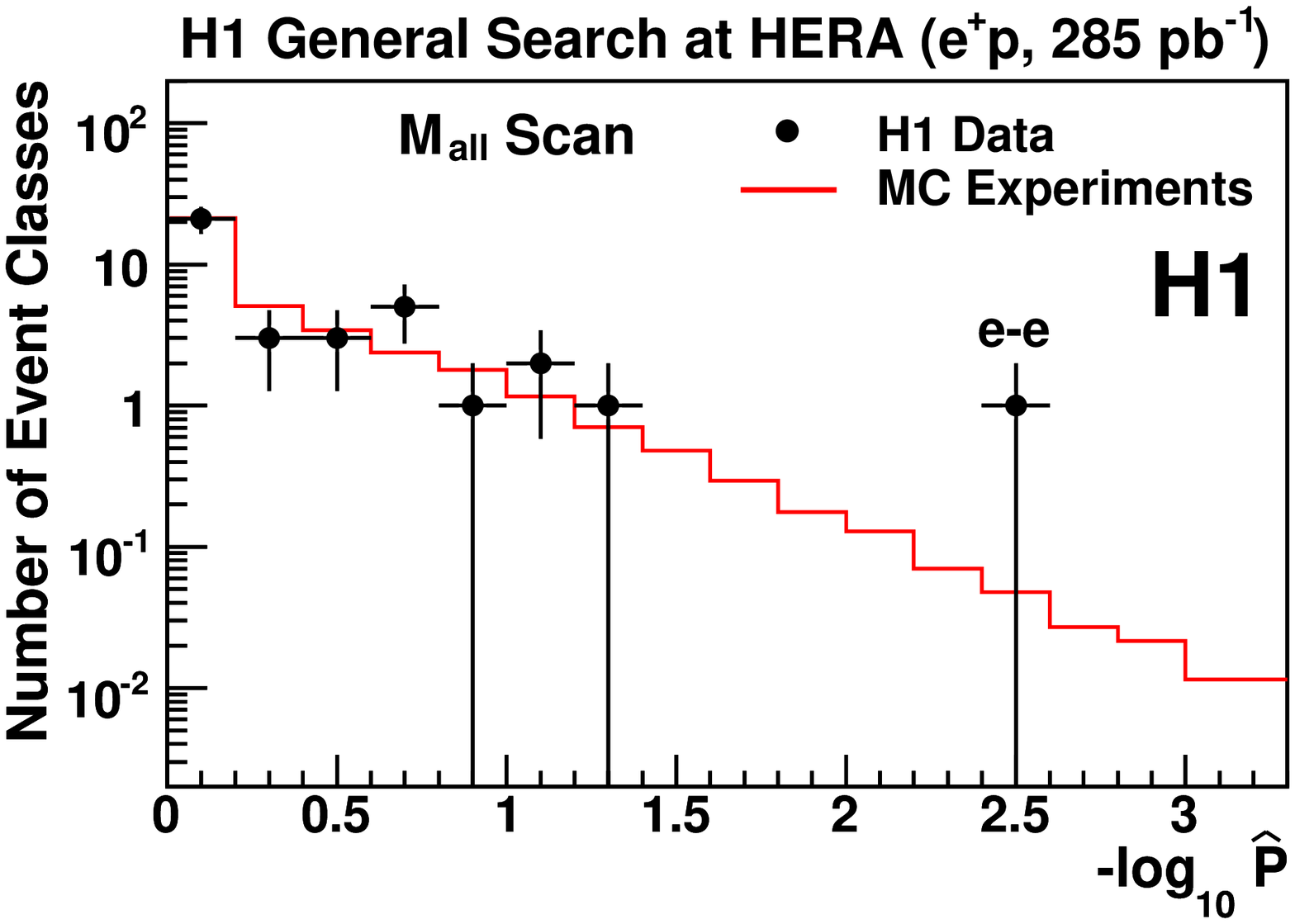}\put(-9.5,30){{(e)}}
  \includegraphics[width=0.49\textwidth]{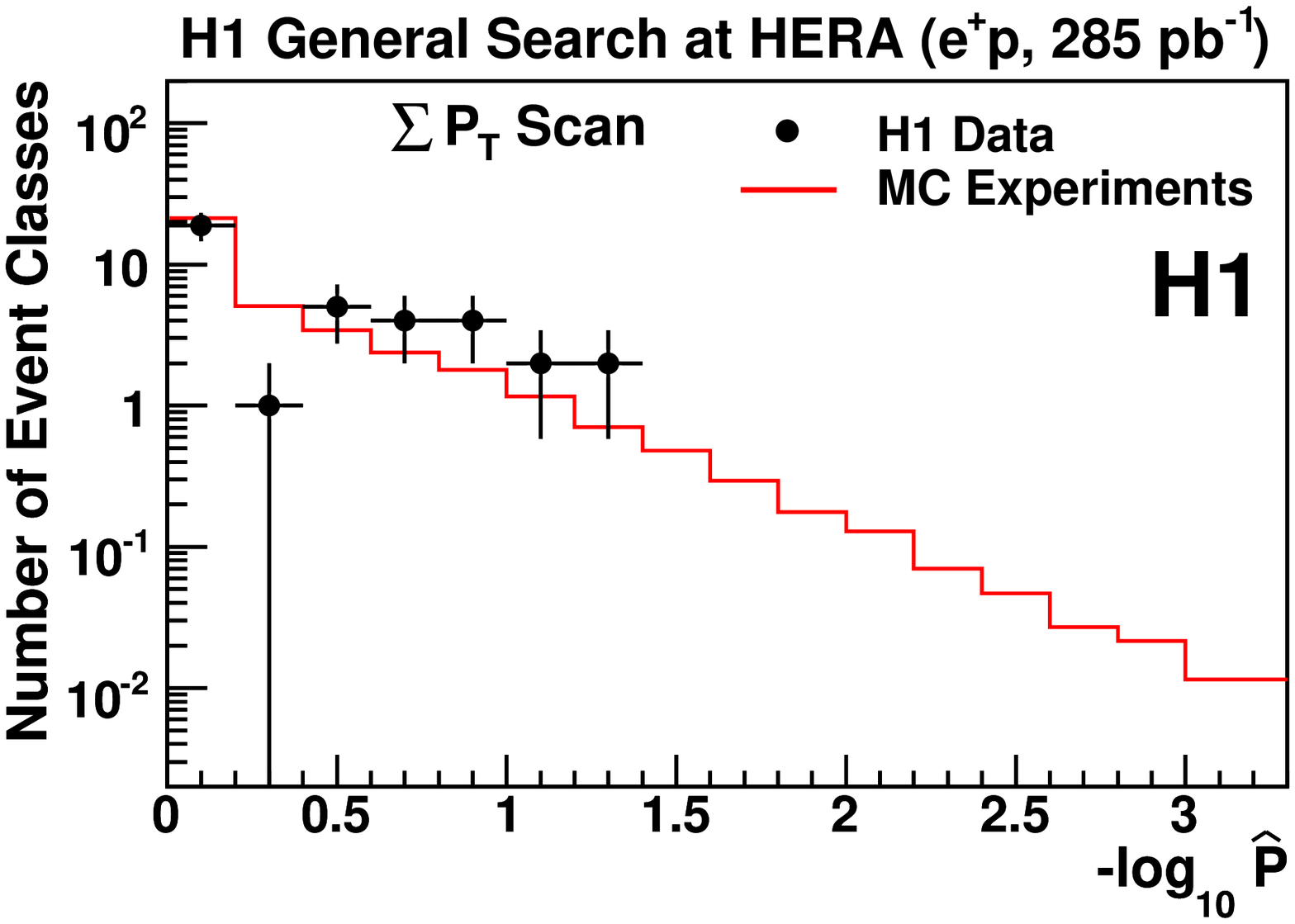}\put(-9.5,30){{(f)}}\\  
    
  \caption{The --$\log_{10}{\hat{P}}$ values for the data event classes and the 
    expected distribution from MC experiments as derived with the search algorithm by investigating
    the $\Mall$ distributions (left column) and $\SPT$ distributions (right column). The results of the scan is presented for all data (a and b), and separately for  $e^-p$ (c and d) and $e^+ p$ (e and f) data.  
    } 
  \label{fig:scan_stat}
\end{figure}

%%%%%%%%%%%%%%
%
The $\hat{P}$ values observed in the data in all event classes are compared in figure~\ref{fig:scan_stat} to the distribution of $\hat{P}$ obtained from a large set of MC experiments. 
The comparison is presented for the scans of the $\Mall$ and $\SPT$ distributions for all $e^\pm p$ data and also separately for $e^-p$ and $e^+p$ data.
The distribution of $\hat{P}$  values measured in the data is in agreement with the expectation from MC experiments.
Using all $e^\pm p$ data, a lowest $\hat{P}$ value of $0.0044$ is found in the \ejj~event class in a region at high transverse momenta, $175 < \SPT < 200$~GeV, where $27$ events are observed for an expectation of $11.6 \pm 1.2$.
In $e^-p$ data, the lowest $\hat{P}$ value is $0.0071$ and corresponds to the \eee~event class where one data event is observed compared to a low SM expectation.
The most significant deviation from SM predictions is measured in $e^+p$ collisions in the \ee~event class with $\hat{P}=0.0035$. In the corresponding region ($110 < \Mall < 120$~GeV) five data events are found while $0.43 \pm 0.04$ are expected. 
The global probability to find in the $e^+ p$ data at least one class with a $\hat{P}$ value smaller than observed in the \ee~event class is $12$\% as deduced from MC experiments.

In case of the $\CosTh$ and $\Xlead$ distributions, no significant discrepancy between the data and the SM expectation is found. The lowest $\hat{P}$ value is $0.017$, observed in the $\Xlead$ distribution of the  \jjjnp~event class. 
In event classes where the SM contribution is high ($> 100$ events), the correlation between $\Mall$ or $\SPT$ distributions and $\CosTh$ and $\Xlead$ is further exploited.
The variables $\CosTh$ and $\Xlead$ are used to select events in a phase space region where the SM contribution is reduced and exotic event topologies may be favoured. 
Events where the leading body is emitted in the forward direction are selected by requiring $\CosTh > 0$.
The variable $\Xlead$ is used in three bodies event classes to select topologies corresponding to a sequential resonance decay by requiring $0.75 < \Xlead <0.9$, as deduced from the study of different MC samples of exotic processes.
After a cut on these variables, an overall good agreement between the data and the SM is still observed in $\Mall$ and $\SPT$ distributions. The complete search procedure and statistical analysis is applied to these distributions, the results of which are summarised in table~\ref{tab:scan_red}. 
No significant deviation is observed in the reduced event samples.

The full analysis is also performed at lower and higher transverse momenta by changing the minimum $P_T$ of particles to $P_T > 15$~GeV and $P_T > 40$~GeV, respectively.
A good overall agreement with the SM is also observed with these cuts.
With a cut $P_T > 15$~GeV, all spectra are well described by the MC, including the multi--jet event classes. The lowest $\hat{P}$ value is $0.01$, observed in the \ejj~event class.
When raising the $P_T$ threshold to $40$~GeV, mainly event classes containing jets remain populated and the largest deviation is observed in the \ejjj~class with $\hat{P}= 0.01$.

%%%%%%%%%%%%%%%%%%%%%%%%%%%%%%%%%%%%%%%%%%%%%%%%%%%%%%%%%%%%%
\section{Conclusion}
%%%%%%%%%%%%%%%%%%%%%%%%%%%%%%%%%%%%%%%%%%%%%%%%%%%%%%%%%%%%%

The full $e^\pm p$ data sample collected by the H1 experiment at HERA is investigated in a general search for deviations from the SM prediction at high transverse 
momenta. 
This analysis encompasses all event topologies involving isolated electrons, photons, 
muons, neutrinos and jets with transverse momenta above $20$~GeV.
Data events are found in $27$ different final states and events with up to five high $P_T$ particles are observed.
In each event class deviations from the SM are searched for in the invariant mass and sum of transverse momenta distributions using a dedicated algorithm. 
In addition, the final state topologies are also evaluated in terms of angular distributions and energy sharing between final state particles. 
A good agreement with the SM expectation is observed in the phase space covered by this analysis.
The largest deviation is found in the \ee~event class, in $e^+p$ collisions, at high invariant masses and corresponds to a probability of $0.0035$.
The probability to observe a SM fluctuation with that significance or higher for at least one event class is $12$\%. 
This comprehensive analysis demonstrates the very good understanding of high
$P_T$ SM phenomena achieved at the HERA collider.
%and is one of the most comprehensive search for new physics beyond the SM performed at a collider experiment on the high energy frontier.

%%%%%%%%%%%%%%%%%%%%%%%%%%%%%%%%%%%%%%%%%%%%%%%%%%%%%%%%%%%%
\section*{Acknowledgements}

We are grateful to the HERA machine group whose outstanding
efforts have made this experiment possible. 
We thank the engineers and technicians for their work in constructing 
and maintaining the H1 detector, our funding agencies for financial 
support, the DESY technical staff for continual assistance and the 
DESY directorate for the hospitality which they extend to the non--DESY 
members of the collaboration.

%%%%%%%%%%%%%%%%%%%%%%%%%%%%%%%%%%%%%%%%%%%%%%%%%%%%%%%%%%%%

%%%%%%%%%%%%%%%%%%%%%%%%%%%%%%%%%%%%%%%%%%%%%%%%%%%%%%%%%%%%%%%%%%%%%%%%%%%%%%
\clearpage

\begin{table}[]
\begin{center}
\footnotesize{
\begin{tabular}{ l c c l l l l}
\multicolumn{7}{c}{{\bf H1 General Search at HERA (\begin{boldmath}$e^\pm p$\end{boldmath}, \begin{boldmath}$463$\end{boldmath} pb\begin{boldmath}$^{-1}$\end{boldmath})}}\\
\hline
 Event class & Data & SM & $\hat{P}_{\SPT}$ & $\hat{P}_{\Mall}$ & $\hat{P}_{\CosTh}$ & $\hat{P}_{\Xlead}$\rule[-6pt]{0pt}{19pt} \\
\hline                                        
\jj & $156724$ & $153278 \pm 27400$ & $0.57$  & $0.33$  & $0.98$  &  \\ 
\ej & $125900$ & $127917 \pm 15490$ & $0.090$ & $0.99$  & $0.40$  &  \\ 
\muj & $21$ & $19.5 \pm 3.0$ 	    & $0.30$  & $0.46$  & $0.024$ &\\ 
\jnp & $11081$ & $11182 \pm 1165$   & $0.33$  & $0.31$  & $0.25$ & \\ 
\enp & $16$ & $21.5 \pm 3.5$        & $0.13$  & $0.084$  & $0.62$ &  \\ 
\ee & $36$ & $40.0 \pm 3.7$           & $0.35$  & $0.041$  & $0.52$ &  \\ 
\emu & $19$ & $21.0 \pm 2.1$        & $0.46$  & $0.83$   & $0.81$ &   \\ 
\mumu & $18$ & $17.5 \pm 3.0$ 	    & $0.31$  & $0.50$   & $0.88$ &    \\ 
\jpho & $563$ & $538 \pm 86$ 	    & $0.31$  & $0.21$   & $0.77$ &     \\ 
\epho & $619$ & $648 \pm 62$ 	    & $0.93$  & $0.99$   & $0.10$ &    \\ 
\mupho & $0$ & ~~$0.22 \pm 0.04$ 	    &  $1$ & $1$ & $1$ &    \\ 
\nppho & $4$ & ~~$9.6 \pm 2.8$ 	    & $0.076$ & $0.33$   & $0.22$ &   \\ 
\phopho & $1$ & ~~$1.1 \pm 0.6$ 	    & $0.66$  & $0.35$ & $0.11$ &  \\ 
\hline
\jjj & $2581$ & $2520 \pm 725$ 	    & $0.54$  & $0.65$ &  & $0.18$  \\ 
\ejj & $1394$ & $1387 \pm 270$ 	    & $0.0044$ & $0.70$ &  & $0.28$\\ 
\mujj & $1$ & ~~$0.46 \pm 0.18$ 	    & $0.12$  & $0.072$  &  & $0.99$\\ 
\jjnp & $355$ & $338 \pm 62$ 	    & $0.80$  & $0.48$ & &$0.62$\\ 
\eej & $0$ & ~~$0.31 \pm 0.04$ 	    &  $1$  &  $1$ &   & $1$\\ 
\eenp & $0$ & ~~$0.06 \pm 0.01$ 	    &  $1$  & $1$ &   &   $1$\\ 
\eee & $1$ & ~~$0.22 \pm 0.04$ 	    & $0.15$  & $0.031$ &   & $0.14$ \\ 
\mumuj & $0$ & ~~$0.16 \pm 0.03$ 	    &  $1$  & $1$ &  & $1$ \\ 
\emumu & $0$ & ~~$0.37 \pm 0.07$ 	    &  $1$  & $1$ & & $1$ \\ 
\mumunp & $0$ & ~~$0.010 \pm 0.005$ 	    &  $1$  & $1$ &  & $1$ \\ 
\emuj & $0$ & ~~$0.16 \pm 0.04$ 	    &  $1$  & $1$ &  & $1$ \\ 
\ejnp & $4$ & ~~$3.2 \pm 0.5$ 	    & $0.24$  & $0.57$ &   & $0.095$ \\ 
\mujnp & $5$ & ~~$2.8 \pm 0.5$ 	    & $0.27$  & $0.30$ &   &   $0.35$ \\ 
\emunp & $0$ & ~~$0.05 \pm 0.01$ 	    &  $1$  & $1$  &   & $1$ \\ 
\jjpho & $5$ & ~~$6.7 \pm 1.3$ 	    & $0.41$  & $0.25$ &   &  $0.91$ \\ 
\ejpho & $12$ & $19.4 \pm 4.0$ 	    & $0.31$  & $0.28$ &   &  $0.53$ \\ 
\jnppho & $1$ & ~~$4.5 \pm 1.5$ 	    & $0.35$  & $0.62$ &   &  $0.47$ \\ 
\hline
\ejjj & $19$ & ~~~$22 \pm 6.5$ 	    & $0.84$  & $0.80$ &   & $0.14$\\ 
\jjjnp & $7$ & ~~$5.2 \pm 1.4$ 	    & $0.47$  & $0.39$ &  & $0.017$ \\ 
\jjnppho & $0$ & ~~$0.16 \pm 0.07$ 	    &  $1$  &  $1$ &    &  $1$ \\ 
\ejjnp & $0$ & ~~$0.15 \pm 0.09$ 	    &  $1$  &  $1$ &    &  $1$ \\ 
\ejjpho & $0$ & ~~$0.22 \pm 0.07$ 	    &  $1$  &  $1$ &    &  $1$ \\ 
\eejnp & $0$ & ~~$0.10 \pm 0.06$ 	    &   $1$ &  $1$ &    &  $1$ \\ 
\emujnp & $0$ & ~~$0.08 \pm 0.05$ 	    &  $1$  &  $1$ &    &  $1$ \\ 
\jjjj & $40$ & ~~$33 \pm 13$ 	    &    &   &    & \\ 
\hline
\ejjjj & $1$ & ~~$0.13 \pm 0.06$ 	    &   &   &    &  \\ 
\jjjjnp & $1$ & ~~$0.05 \pm 0.02$     &   &    &   & \\ 
\jjjjj & $0$ & ~~$0.14 \pm 0.09$     &   &     &    & \\ 
\hline
\end{tabular}
}
\end{center}
\caption{Observed and predicted event yields for all event classes with observed data events or a SM expectation greater than $0.01$ for all $e^{\pm}p$ data.
Each event class is labeled with the leading body listed first.
  The errors on the predictions include model uncertainties and experimental systematic errors added in quadrature.
  The $\hat{P}$ values obtained in the scan of $\SPT$,  $\Mall$, $\CosTh$ and $\Xlead$ distributions are also given.}
\label{tab:yields}
\end{table}

\begin{table}[]
\begin{center}
\begin{tabular}{ l l c c l l }
\multicolumn{6}{c}{{\bf H1 General Search at HERA (\begin{boldmath}$e^\pm p$\end{boldmath}, \begin{boldmath}$463$\end{boldmath} pb\begin{boldmath}$^{-1}$\end{boldmath})}}\\
\hline
Event class & Selection & Data & SM & $\hat{P}_{\SPT}$ & $\hat{P}_{\Mall}$ \rule[-6pt]{0pt}{19pt}\\
\hline                                        
\jj & $\CosTh > 0$  & $83155$ & ~~$82800 \pm 15610$ & $0.46$   & $0.44$ \\ 
\ej & $\CosTh > 0$  & $6532$ & $6603 \pm 783$ & $0.23$  & $0.033$ \\ 
\jnp & $\CosTh > 0$  & $2177$ & $2076 \pm 240$    & $0.61$   & $0.75$ \\ 
\jpho & $\CosTh > 0$ & $123$ & $118 \pm 20$ 	      & $0.15$   & $0.016$\\ 
\epho & $\CosTh > 0$ & $227$ & $260 \pm 25$ 	      & $0.12$   & $0.19$ \\ 
\jjj  & $\CosTh > 0$ & 	$1359$  & $1218 \pm 340$  & $0.36$   &  $0.63$\\ 
\ejj  & $\CosTh > 0$ & $65$   &  ~~$74 \pm 13$ & $0.75$ & $0.37$\\ 
\jjnp & $\CosTh > 0$ &  $58$  & ~~$53 \pm 12$  & $0.62$ & $0.26$\\ 
\hline
\jjj  &  $0.75 < \Xlead < 0.9$ &  $1672$     & $1658 \pm 482$  &   $0.096$ &  $0.40$  \\ 
\ejj  &  $0.75 < \Xlead < 0.9$ &  $419$      &  $419 \pm 81$ &  $0.018$  &  $0.07$    \\ 
\jjnp &  $0.75 < \Xlead < 0.9$ &  $133$      &  $109 \pm 22$ & $0.26$  &  $0.19$     \\ 

\hline
\end{tabular}
\end{center}
\caption{Observed and predicted event yields for considered event classes after a cut on the topological variables.
Each event class is labeled with the leading body listed first.
  The errors on the predictions include model uncertainties and experimental systematic errors added in quadrature.
  The $\hat{P}$ values obtained in the scan of $\SPT$ and $\Mall$ distributions are indicated in the last two columns.}
\label{tab:scan_red}
\end{table}

%=========================================================================
\vfill
\newpage

\clearpage

\clearpage

\clearpage

\newpage
\clearpage

\end{document}